\documentclass[12pt]{article}
\usepackage{amsmath,amssymb,bbm,stmaryrd}

\newcommand{\ft}[2]{{\textstyle\frac{#1}{#2}}}

\def\slashchar#1{\setbox0=\hbox{$#1$}           
   \dimen0=\wd0                                 
   \setbox1=\hbox{/} \dimen1=\wd1               
   \ifdim\dimen0>\dimen1                        
      \rlap{\hbox to \dimen0{\hfil/\hfil}}      
      #1                                        
   \else                                        
      \rlap{\hbox to \dimen1{\hfil$#1$\hfil}}   
      /                                         
   \fi}

\def\diag{\text{diag}}
\def\Re{\text{Re}}
\def\Im{\text{Im}}

\def\mua{\mu}
\def\mub{\nu}
\def\muc{\rho}
\def\mud{\lambda}
\def\mue{\sigma}

\def\sMa{{\cal M}}
\def\sMb{{\cal N}}
\def\sMc{{\cal P}}
\def\sMd{{\cal Q}}
\def\sMe{{\cal R}}
\def\sMf{{\cal S}}
\def\sMg{{\cal T}}

\def\La{\Lambda}
\def\Lb{\Sigma}
\def\Lc{\Gamma}
\def\Ld{\Delta}
\def\Le{\Xi}
\def\Lf{\Phi}

\def\Le{\Upsilon}
\def\Ma{M}
\def\Mb{N}
\def\Mc{P}
\def\Md{Q}
\def\Me{R}
\def\Mf{S}
\def\Mg{T}
\def\Mh{U}
\def\Mi{V}
\def\ja{i}
\def\jb{j}
\def\jc{k}
\def\jd{l}
\def\je{m}
\def\jf{n}
\def\jg{p}
\def\jh{q}
\def\ji{r}
\def\jj{s}
\def\jk{t}
\def\jl{u}
\def\ya{m}
\def\yb{n}
\def\yc{o}
\def\yd{p}
\def\ye{q}
\def\yf{r}
\def\yg{s}
\def\yh{t}
\def\xa{a}

\def\aa{\alpha}
\def\ab{\beta}
\def\ac{\gamma}
\def\ad{\delta}

\def\Na{{\hat 0}}
\def\Nv{0}
\def\cMa{\sMa}
\def\cMb{\sMb}
\def\cMc{\sMc}
\def\cMd{\sMd}
\def\cMe{\sMe}
\def\cMf{\sMf}
\def\cMg{\sMg}
\def\La{\Lambda}
\def\Lb{\Gamma}
\def\Lc{\Sigma}
\def\Ld{\Psi}
\def\Le{\Delta}
\def\Lf{\Xi}

\begin{document}

\begin{titlepage}
\begin{center}

\hfill {\tt DESY/06-009}\\
\hfill {\tt ZMP-HH/06-01}

\vskip 1.5cm 
\begin{center}
{\Large {\bf Gauged N$=$4 supergravities}}
\end{center}

\vskip 1.0cm

{\bf Jonas Sch\"on and Martin Weidner} \\

\vskip 22pt

{\em II. Institut f\"ur Theoretische Physik\\[-.6ex]
Universit\"at Hamburg\\[-.6ex]
Luruper Chaussee 149\\[-.6ex]
D-22761 Hamburg, Germany}

\vskip 8pt
and
\vskip 8pt

{\em Zentrum f\"ur Mathematische Physik\\[-.6ex]
Universit\"at Hamburg\\[-.6ex]
Bundesstrasse 55\\[-.6ex]
D-20146 Hamburg, Germany}\\

\vskip 15pt

{{\tt jonas.schoen@desy.de, martin.weidner@desy.de}} \\

\vskip 0.8cm

\end{center}

\vskip 5pt

\begin{center} {\bf ABSTRACT}\\[3ex]

\begin{minipage}{13cm}
\small

We present the gauged $N=4$ (half-maximal) supergravities in four and five spacetime dimensions coupled to an arbitrary number
of vector multiplets. The gaugings
are parameterized by a set of appropriately constrained constant tensors, which transform covariantly under the
global symmetry groups ${\rm SL}(2) \times {\rm SO}(6,n)$ and ${\rm SO}(1,1) \times {\rm SO}(5,n)$, respectively.
In terms of these tensors the universal Lagrangian and the Killing Spinor equations are given.
The known gaugings, in particular those originating from flux compactifications, are incorporated in the formulation, but also
new classes of gaugings are found.
Finally, we present the embedding chain of the five dimensional into the four dimensional into the three dimensional gaugings,
thereby showing how the deformation parameters organize under the respectively larger duality groups.

\end{minipage}
\end{center}
\noindent

%

\end{titlepage}

\tableofcontents

\newpage

\section{Introduction}
\setcounter{equation}{0}

The first examples of $N=4$ supergravities in four spacetime dimensions were constructed in the second half of 
the seventies \cite{Das:1977uy, Cremmer:1977tc,Cremmer:1977tt,Freedman:1978ra}
and within the following decade the coupling of vector multiplets
to these theories and some of their gaugings were worked out 
\cite{Gates:1982ct,deRoo:1984gd,deRoo:1985jh,deRoo:1986yw,Bergshoeff:1985ms}. 
In $N=4$ the gaugings are the only known deformations of the theory that are compatible with supersymmetry.
They are induced by minimal couplings of vector fields to isometry generators, but supersymmetry
requires various additional couplings and in particular the emergence of a scalar potential, thus giving the
possibility of ground states with non-vanishing cosmological constant.
So far, however, no stable de Sitter ground state has been found in these theories \cite{deRoo:2003rm}.

From a string theory perspective the $N=4$ theories result from orientifold compactifications of IIB supergravity
\cite{Frey:2002hf,Kachru:2002he}.
In this picture
part of the deformation parameters of the gauging correspond to fluxes or additional branes on the background
\cite{D'Auria:2002tc,D'Auria:2003jk,Angelantonj:2003rq,Angelantonj:2003up,Berg:2003ri}. But so far not all known gaugings
could be identified in this way.
Lower $N$ theories can be obtained by truncation of the $N=4$ supergravities, for example certain relevant $N=1$
K\"ahler potentials can be computed from the $N=4$ scalar potential \cite{Derendinger:2004jn,Derendinger:2005ph,Derendinger:2006ed}.

By incorporating all possible gauged $N=4$ supergravities in a universal formulation in this paper
we hope to illuminate the interrelation of the different theories but also to pave the way for a future analysis
of particular gaugings. The gaugings are parameterized by an embedding tensor which can be treated as a group
theoretical object and is subject to a set of consistency constraints. This method
was successfully used to work out the general gaugings of maximal
supergravities for various spacetime dimensions \cite{Nicolai:2000sc,deWit:2002vt,deWit:2004nw,Samtleben:2005bp}.
For an even number of spacetime dimensions there are subtleties that seem to hamper the universal description.
For example in $D=4$ magnetic vector fields are usually introduced on-shell via the equations of motion,
while for a general gauging they may possibly occur as gauge fields in the covariant derivative already 
at the level of the Lagrangian.
Closely related to this problem is the fact that in $D=4$
the global symmetry group of a supergravity theory is generically only realized on-shell
since it involves duality rotations between the electric and magnetic vector fields \cite{Gaillard:1981rj,deWit:2001pz}. 
These issues were resolved in \cite{deWit:2005ub}, where for a general four dimensional theory it was explained
how to consistently couple electric and magnetic vector gauge fields together with two-form tensor gauge fields for 
a general gauging. Here we apply this method to the case of gauged $N=4$ supergravities.

In $D=4$ the global symmetry group of the ungauged theory is $G={\rm SL}(2) \times {\rm SO}(6,n)$, 
where $n$ denotes the number of vector multiplets. This group also organizes the
gaugings since the deformation parameters $f_{\aa\Ma\Mb\Mc}$ and $\xi_{\aa\Ma}$ are tensors under $G$
(they are explicitly defined below).
These tensors are the irreducible components of the embedding tensor.
In terms of them the bosonic Lagrangian and the Killing spinor equations are presented,
the consistency constraints which they have to satisfy are explained and solutions to these constraints
are discussed. In particular the ${\rm SU}(1,1)$ phases that were introduced by de Roo and Wagemans 
to find ground states with non-vanishing cosmological constant
\cite{deRoo:1985jh,deRoo:1986yw,Wagemans:1990mv} are identified as parameters incorporated in $f_{\aa\Ma\Mb\Mc}$.
In the same manner the parameters that correspond to three-form fluxes in compactifications
from IIB supergravity \cite{D'Auria:2002tc,D'Auria:2003jk,Angelantonj:2003rq,Angelantonj:2003up} are identified.
Also the gaugings that originate from Scherk-Schwarz reduction from $D=5$ are included in our formulation
\cite{Villadoro:2004ci}. 
In addition, there are various other gaugings that have not yet been discussed in the literature, in particular all gaugings
with both $f_{\aa\Ma\Mb\Mc}$ and $\xi_{\aa\Ma}$ non-zero are novel.

Analogous to the four dimensional case
the general five dimensional gauged $N=4$ supergravity\footnote{We denote
by $N=4$ the half-maximal supergravity, although in five spacetime dimensions this theory is sometimes referred to as $N=2$.}
is worked out by applying the ideas of \cite{deWit:2004nw},
where the corresponding gauged maximal supergravity was presented.
In $D=5$ the irreducible components of the embedding tensor are tensors $f_{\Ma\Mb\Mc}$, $\xi_{\Ma\Mb}$ and $\xi_\Ma$,
which are tensors under the global symmetry group ${\rm SO}(1,1) \times {\rm SO}(5,n)$.
The first account of the ungauged $N=4$ supergravity in $D=5$ was given in \cite{Awada:1985ep},
where also the first gauging of the theory was already considered.
Those gaugings where the gauge group is a product of a semi-simple and an Abelian factor
were already presented in \cite{Dall'Agata:2001vb}, examples of this type were already known for a while
\cite{Romans:1985ps}. Also some non-semi-simple gaugings
were already constructed \cite{Villadoro:2004ci}. Our presentation incorporates all these known gaugings
and also includes new ones.

In former descriptions of $D=5$ gauged supergravities the vector fields that are not needed as gauge fields
were dualized into two-form fields to make the theory consistent
\cite{Dall'Agata:2001vb,Romans:1985ps,Pernici:1985ju,Gunaydin:1985cu,Gunaydin:1999zx,Andrianopoli:2000fi,Bergshoeff:2004kh}.
This makes the field content of the theory dependent on the particular gauging and makes it difficult to formulate
the general gauged theory in a covariant way. It was shown in \cite{deWit:2004nw} that one can deal with this issue by introducing
both the vector fields and all their dual two-form fields as off-shell degrees of freedom
and couple them via a topological term such that
their duality equation results from the equations of motion.
The same concept is used here to describe the general five dimensional gauged theory.

The gauged $N=4$ supergravities in five dimensions are naturally embedded into the four dimensional ones by dimensional reduction
and we make this relation explicit within this paper. Noteworthy, the five dimensional gaugings are parameterized in
terms of three tensors $f_{\Ma\Mb\Mc}$, $\xi_{\Ma\Mb}$ and $\xi_{\Ma}$ while the four dimensional ones are parameterized
in terms of two tensors $f_{\aa\Ma\Mb\Mc}$ and $\xi_{\aa\Ma}$ only. Thus with decreasing spacetime dimension one finds not
only a larger duality group but also a more uniform
description of the deformations. This is the typical picture of dualities in string theory where dimensional
reduction relates theories with different higher-dimensional origin.

The paper is organized as follows. In section \ref{sec:D4} we present the general four dimensional theory. We give its bosonic
Lagrangian and its Killing spinor equations, discuss the consistency constraints on the deformation parameters,
and describe examples of gaugings, including those known from the literature.
In section \ref{sec:D5} the five dimensional theories are discussed analogously.
Eventually, having both general gauged theories at hand, their embedding induced by a circle reduction
is given. For completeness, we sketch the analogous embedding of the $D=4$ into the $D=3$ gaugings in the appendix.

\section{Gauged $N=4$ supergravities in $D=4$}
\setcounter{equation}{0}
\label{sec:D4}

The gaugings of $N=4$ supergravity in four spacetime dimensions are parameterized by two real constant 
tensors $f_{\aa\Ma\Mb\Mc}$
and $\xi_{\aa\Ma}$. These are tensors under the global on-shell symmetry group ${\rm SL}(2) \times {\rm SO}(6,n)$,
and $\aa=1,2$ and $\Ma=1,\ldots,6+n$ are the respective vector indices.
In the following section the Lagrangian of the theory is given in terms of these tensors.
However, $f_{\aa\Ma\Mb\Mc}$ and $\xi_{\aa\Ma}$ can not be chosen arbitrarily,
the consistency conditions that they have to obey are discussed in section \ref{sec:D4con}.

\subsection{Lagrangian and field equations}
\label{sec:D4lag}

The $N=4$ supergravity multiplet contains as bosonic degrees of freedom the metric, six massless vectors and two real
massless scalars.
The corresponding supergravity theory has a global ${\rm SL}(2) \times {\rm SO}(6)$ symmetry \cite{Cremmer:1977tt}
which is realized only on-shell.
The scalar fields constitute an ${\rm SL}(2)/{\rm SO}(2)$ coset\footnote{
In the literature the symmetry group is usually denoted by ${\rm SU}(1,1)$, however, we prefer to treat it as ${\rm SL}(2)$ 
which is of course the same group but with different conventions concerning its fundamental representation.}.
Coupling this theory to $n$ vector multiplets, each containing one
vector and six real scalars, yields an $N=4$ supergravity with global on-shell symmetry group $G={\rm SL}(2) \times {\rm SO}(6,n)$
\cite{deRoo:1984gd}. This is the theory whose deformations we want to study here for arbitrary $n \in \mathbb{N}$. 

For the vector fields of the theory one can choose a symplectic frame such that the subgroup
${\rm SO}(1,1) \times {\rm SO}(6,n)$ of $G$ is realized off-shell.
The electric vector fields ${A_\mua}^{\Ma+}$ ($\Ma = 1, \ldots, 6+n$) then form a vector under ${\rm SO}(6,n)$ and carry charge $+1$
under ${\rm SO}(1,1)$. Their dual magnetic vector fields ${A_\mua}^{\Ma-}$ form an ${\rm SO}(6,n)$ vector as well but carry
${\rm SO}(1,1)$ charge $-1$. Together they constitute an ${\rm SL}(2)$ vector
${A_\mua}^{\Ma\aa}=({A_\mua}^{\Ma+},{A_\mua}^{\Ma-})$\footnote{
Here and in the following we use indices $\aa,\ab, \ldots = +,-$ for ${\rm SL}(2)$ vectors. The embedding of the off-shell
symmetry group ${\rm SO}(1,1)$ into ${\rm SL}(2)$ defines a basis for these vectors and thus
components $v^\alpha=(v^+,v^-)$ and $v_\alpha=(v_+,v_-)$. For the epsilon tensor $\epsilon_{\aa\ab}$ we use
$\epsilon_{+-}=\epsilon^{+-}=1$ which yields $\epsilon_{\aa\ac} \epsilon^{\ab\ac}=\delta_\aa^\ab$.}.

The scalar fields form the coset space
$G/H$, where $H={\rm SO}(2) \times {\rm SO}(6) \times {\rm SO}(n)$ is the maximal compact subgroup of $G$.
The ${\rm SL}(2)/{\rm SO}(2)$ factor of this coset can 
equivalently be described by a complex number $\tau$ with $\Im(\tau)>0$ or
by a symmetric positive definite matrix $M_{\aa\ab} \in {\rm SL}(2)$.
The relation between these two descriptions is given by
\begin{align}
   M_{\aa\ab} &= \frac 1 {\Im(\tau)} \begin{pmatrix} |\tau|^2 & \Re(\tau) \\ \Re(\tau) & 1  \end{pmatrix}\;, &
   M^{\aa\ab} &= \frac 1 {\Im(\tau)} \begin{pmatrix} 1 & -\Re(\tau) \\ -\Re(\tau) & |\tau|^2 \end{pmatrix}\;, 
\end{align}
where $M^{\aa\ab}$ is the inverse of $M_{\aa\ab}$. The ${\rm SL}(2)$ symmetry action on $M_{\aa\ab}$
\begin{align}
   M &\rightarrow g M g^T \;,&
   g &= \begin{pmatrix} a & b \\ c & d \end{pmatrix} \; \in {\rm SL}(2) \; ,
\end{align}
acts on $\tau$ as a M\"obius transformation $\tau \rightarrow (a \tau + b)/(c \tau + d)$.

The ${\rm SO}(6,n)/{\rm SO}(6)\times{\rm SO}(n)$
factor of the scalar coset is described by coset representatives ${{\cal V}_\Ma}^{\xa}$ and ${{\cal V}_\Ma}^{\ya}$
where $\ya=1,\ldots,6$ and $\xa=1,\ldots,n$ denote ${\rm SO}(6)$ and ${\rm SO}(n)$
vector indices, respectively. The matrix ${\cal V}=({{\cal V}_\Ma}^{\ya},\,{{\cal V}_\Ma}^{\xa})$ 
is an element of ${\rm SO}(6,n)$, i.e.
\begin{align}
   \eta_{\Ma\Mb} &= - {{\cal V}_\Ma}^{\ya} {{\cal V}_\Mb}^{\ya} + {{\cal V}_\Ma}^\xa {{\cal V}_\Mb}^\xa \; ,
   \label{DefEta}
\end{align}
where $\eta_{\Ma\Mb}=\diag(-1, -1, -1, -1, -1, -1, +1, \ldots, +1)$ is the ${\rm SO}(6,n)$ metric.
Global ${\rm SO}(6,n)$ transformations act on ${\cal V}$ from the left while local ${\rm SO}(6)  \times {\rm SO}(n)$ transformations
act from the right
\begin{align}
   {\cal V} \, &\rightarrow \, g {\cal V} h(x) \; , &&
   g \in {\rm SO}(6,n), \quad h(x) \in {\rm SO}(6)  \times {\rm SO}(n) \;.
   \label{CosetSO6n}
\end{align}
Analogous to $M_{\aa\ab}$ this coset space may be parameterized by
a symmetric positive definite scalar metric $M={\cal V}{\cal V}^T$,
explicitly given by
\begin{align}
   M_{\Ma\Mb} &= {{\cal V}_\Ma}^\xa {{\cal V}_\Mb}^\xa + {{\cal V}_\Ma}^{\ya} {{\cal V}_\Mb}^{\ya} \; .
   \label{DefMVV}
\end{align}
Its inverse we denote by $M^{\Ma\Mb}$. Note that each of the matrices $M_{\Ma\Mb}$, ${{\cal V}_\Ma}^{\ya}$
and ${{\cal V}_\Ma}^{\xa}$ alone already parameterizes the ${\rm SO}(6,n)$ part of the scalar coset.

In order to later give the scalar potential
we also need to define the scalar dependent completely antisymmetric tensor
\begin{align}
   M_{\Ma\Mb\Mc\Md\Me\Mf} &= \epsilon_{\ya\yb\yc\yd\ye\yf} \,
                             {{\cal V}_\Ma}^{\ya} {{\cal V}_\Mb}^{\yb} {{\cal V}_\Mc}^{\yc} 
                             {{\cal V}_\Md}^{\yd} {{\cal V}_\Me}^{\ye} {{\cal V}_\Mf}^{\yf}  \; .
   \label{DefM6}			     
\end{align}

The ungauged theory contains the metric, electric vector fields and scalars as free fields in the Lagrangian,
while the dual magnetic vectors and two-form gauge fields are only introduced on-shell (this is the description we choose).
The latter come in the adjoint representation of $G$ and since $G$ has two factors there are also two kinds of
two-form gauge fields, namely $B_{\mua\mub}^{\Ma\Mb}=B_{\mua\mub}^{[\Ma\Mb]}$ and 
$B_{\mua\mub}^{\aa\ab}=B_{\mua\mub}^{(\aa\ab)}=(B_{\mua\mub}^{++},B_{\mua\mub}^{+-},B_{\mua\mub}^{--})$.
For the general description of the gauged theory all these fields appear as free fields in the Lagrangian
\cite{deWit:2005ub}.
For the magnetic vectors this is necessary because they can appear as gauge fields
in the covariant derivative while the two-forms in turn are required in order to consistently couple the vector fields.
Some of the vector fields that are not needed in the gauging become Stueckelberg fields for the two-forms.

Neither the magnetic vector fields ${A_\mua}^{\Ma-}$ nor the two-form gauge fields have a kinetic term
and via their first order equations of motion they eventually turn out to be dual to the electric vector fields ${A_\mua}^{\Ma+}$
and to the scalars, respectively. Thus the number of degrees of freedom remains unchanged as compared to the ungauged theory.

The gauged supergravities are parameterized by two $G$-tensors $\xi_{\aa\Ma}=(\xi_{+\Ma},\xi_{-\Ma})$ and
$f_{\aa\Ma\Mb\Mc}=(f_{+\Ma\Mb\Mc},f_{-\Ma\Mb\Mc})$ with $f_{\aa\Ma\Mb\Mc}=f_{\aa[\Ma\Mb\Mc]}$.
One should think of these tensors as generalized structure constants of the gauge group.
They have to satisfy certain consistency constraints to be introduced later. The following combinations occur regularly
\begin{align}
   \Theta_{\aa\Ma\Mb\Mc} &= f_{\aa\Ma\Mb\Mc} - \xi_{\aa[\Mb} \, \eta_{\Mc]\Ma} \; , \nonumber \\
   {\hat f}_{\aa\Ma\Mb\Mc} &= f_{\aa\Ma\Mb\Mc} - \xi_{\aa[\Ma}  \, \eta_{\Mc]\Mb} - \, \ft 3 2 \, \xi_{\aa\Mb} \eta_{\Ma\Mc} \; .
   \label{DefThetaF}
\end{align}
In addition we use a gauge coupling constant $g$ which is actually dispensable
by rescaling $f_{\aa\Ma\Mb\Mc} \, \rightarrow \, g^{-1} \, f_{\aa\Ma\Mb\Mc}$ and
$\xi_{\aa\Ma} \, \rightarrow \, g^{-1} \, \xi_{\aa\Ma}$. Nevertheless it is convenient
to use $g$ to keep track of the order in the gauge coupling.

We can now present the bosonic Lagrangian of the general gauged theory\footnote{
Our space-time metric has signature $(-,+,+,+)$ and the Levi-Civita is a proper space-time tensor, i.e.
$\epsilon^{0123}=e^{-1}$, $\epsilon_{0123}=-e$.}
\begin{align}
   {\cal L}_{\text{bos}} &= {\cal L}_{\text{kin}} + {\cal L}_{\text{top}} + {\cal L}_{\text{pot}} \; .
   \label{LagBosD4}
\end{align}
It consists of a kinetic term
\begin{align} 
   e^{-1} {\cal L}_{\text{kin}} &= \ft 1 2 \, R 
                    + \ft 1 {16} \, (D_\mua M_{\Ma\Mb}) (D^\mua M^{\Ma\Mb})
     - \, \frac 1 {4 \, \Im(\tau)^2}  (D_\mua \tau) (D^\mua \tau^*) 
     \nonumber \\[1ex] 
     &  \qquad
     - \, \ft 1 4 \, \Im(\tau) \, M_{\Ma\Mb} {\cal H}_{\mua\mub}{}^{\Ma+} {\cal H}^{\mua\mub\Mb+}
               + \, \ft 1 8 \, \Re(\tau) \, \eta_{\Ma\Mb} \, \epsilon^{\mua\mub\muc\mud} 
    {{\cal H}_{\mua\mub}}^{\Ma+} {{\cal H}_{\muc\mud}}^{\Mb+}   \;,
\end{align}
a topological term for the vector and tensor gauge fields \cite{deWit:2005ub}
\begin{align}
   e^{-1} {\cal L}_{\text{top}} &= - \, \frac g 2 \, \epsilon^{\mua\mub\muc\mud}  \nonumber \\ & \quad
          \bigg\{ 
          \xi_{+\Ma} \eta_{\Mb\Mc} A_\mua^{\Ma-} A_\mub^{\Mb+} \partial_\muc A_\mud^{\Mc+} 
     - \left( \hat f_{-\Ma\Mb\Mc} + 2 \, \xi_{-\Mb} \eta_{\Ma\Mc} \right) A_\mua^{\Ma-} A_\mub^{\Mb+} \partial_\muc A_\mud^{\Mc-} 
	\nonumber \\ & \quad        
     - \, \frac g 4 \, \hat f{}_{\aa\Ma\Mb\Me} \hat f{}_{\ab\Mc\Md}{}^\Me A_\mua^{\Ma\aa} A_\mub^{\Mb+} A_\muc^{\Mc\ab} A_\mud^{\Md-} 
     + \, \frac g {16} \, \Theta_{+\Ma\Mb\Mc} {{\Theta_{-}}^{\Ma}}_{\Md\Me} B_{\mua\mub}^{\Mb\Mc} B_{\muc\mud}^{\Md\Me} 
	\nonumber \\ & \quad  
       - \ft 1 4 \left( \Theta_{-\Ma\Mb\Mc} B_{\mua\mub}^{\Mb\Mc} 
		                     + \xi_{-\Ma} B_{\mua\mub}^{+-} + \xi_{+\Ma} B_{\mua\mub}^{++} \right)
	 \big( 2 \partial_\muc A_\mud^{\Ma-} - g {\hat f}{}_{\aa\Md\Me}{}^\Ma A_\muc^{\Md\aa} A_\mud^{\Me-} \big)
	  \bigg\} \, ,
\end{align}
and a scalar potential
\begin{align}
   e^{-1} {\cal L}_{\text{pot}} &= - g^2 V   
     \nonumber \\ & = - \frac{g^2} {16} \bigg\{
      f_{\aa\Ma\Mb\Mc} f_{\ab\Md\Me\Mf} M^{\aa\ab} \Big[
           \ft 1 3 \, M^{\Ma\Md} M^{\Mb\Me} M^{\Mc\Mf} 
        + ( \ft 2 3 \, \eta^{\Ma\Md} -  M^{\Ma\Md} ) \eta^{\Mb\Me} \eta^{\Mc\Mf}   \Big]
	\nonumber \\ & \qquad \qquad
	        - \ft 4 9 \, f_{\aa\Ma\Mb\Mc} f_{\ab\Md\Me\Mf} \epsilon^{\aa\ab} M^{\Ma\Mb\Mc\Md\Me\Mf}
                + 3 \, \xi_\aa^\Ma \xi_\ab^\Mb  M^{\aa\ab} M_{\Ma\Mb}  \bigg\} \; .
   \label{VD4}		
\end{align}
The covariant derivative $D_\mua$ appearing in ${\cal L}_{\text{kin}}$ acts on objects
in an arbitrary representation of $G={\rm SL}(2) \times {\rm SO}(6,n)$ as
\begin{align}
    D_\mua      &= \nabla_\mua 
            - g \, A_\mua{}^{\Ma\aa} {\Theta_{\aa\Ma}}^{\Mb\Mc} t_{\Mb\Mc}
            + g \, A_\mua{}^{\Ma(\aa} \epsilon^{\ab)\ac} \xi_{\ac\Ma} t_{\aa\ab} \; ,
   \label{CovDivD4}      
\end{align}
where $\nabla_\mua$ contains the spin-connection and $t_{\Mb\Mc}$ and $t_{\aa\ab}$ are
the generators of the global symmetry group\footnote{
In the vector representation the symmetry generators have the form
$(t_{\Ma\Mb})_\Mc{}^\Md = \delta^\Md_{[\Ma} \eta^{\phantom{\Md}}_{\Mb]\Mc}$ and
$(t_{\aa\ab})_\ac{}^\ad = \delta^\ad_{(\aa} \epsilon^{\phantom{\ad}}_{\ab) \ac}$, respectively.
}. Explicitly one finds for the scalar fields
\begin{align}
   D_\mua M_{\aa\ab} &= \partial_\mua M_{\aa\ab} + g A_\mua^{\Ma\ac} \xi_{(\aa\Ma} M_{\ab)\ac}
                              - g A_\mua^{\Ma\ad} \xi_{\epsilon\Ma} \epsilon_{\ad(\aa} \epsilon^{\epsilon\ac} M_{\ab)\ac} \; , \nonumber \\
   D_\mua M_{\Ma\Mb} &= \partial_\mua M_{\Ma\Mb} + 2 g A_\mua{}^{\Mc\aa} {\Theta_{\aa\Mc(\Ma}}^{\Md} M_{\Mb)\Md}   \; .
\end{align}
Note that $\Im(\tau)^{-2} (D_\mua \tau) (D^\mua \tau^*) = - \ft 1 2 (D_\mua M_{\aa\ab}) (D^\mua M^{\aa\ab})$,
i.e. the kinetic term for $\tau$ can equivalently be expressed in terms of $M_{\aa\ab}$.

The full covariant field strengths of the electric and magnetic vector fields are given by\footnote{
Note that the indices $+$ and $-$ on the vector fields and on their field strengths
distinguish the electric ones from the magnetic ones and thus do not indicate complex self-dual combinations
of the field strengths as
is common in the literature. We hope note to confuse the reader with that notation.}
\begin{align}
   {\cal H}_{\mua\mub}^{\Ma+} &= 2 \partial_{[\mua} {A_{\mub]}}^{\Ma+} 
            - g \, \hat f{}_{\aa\Mb\Mc}{}^\Ma {A_{[\mua}}^{\Mb\aa} {A_{\mub]}}^{\Mc+} 
	       \nonumber \\ & \qquad \qquad
	    + \frac g 2 \, {{\Theta_{-}}^{\Ma}}_{\Mb\Mc} B_{\mua\mub}^{\Mb\Mc}
	    + \frac g 2 \, {\xi_+}^\Ma B_{\mua\mub}^{++} + \frac g 2 {\xi_{-}}^\Ma B_{\mua\mub}^{+-} \; ,
   \nonumber \\	    
   {\cal H}_{\mua\mub}^{\Ma-} &= 2 \partial_{[\mua} {A_{\mub]}}^{\Ma-} 
            - g \, \hat f{}_{\aa\Mb\Mc}{}^\Ma {A_{[\mua}}^{\Mb\aa} {A_{\mub]}}^{\Mc-} 
	       \nonumber \\ & \qquad \qquad
	    - \frac g 2 \, {{\Theta_{+}}^{\Ma}}_{\Mb\Mc} B_{\mua\mub}^{\Mb\Mc}
	    + \frac g 2 \, {\xi_-}^\Ma B_{\mua\mub}^{--} + \frac g 2 {\xi_{+}}^\Ma B_{\mua\mub}^{+-} \; .
    \label{FieldStrD4}	    
\end{align}
Only ${\cal H}_{\mua\mub}^{\Ma+}$ enters the Lagrangian, but ${\cal H}_{\mua\mub}^{\Ma-}$ appears in the equations of
motion. To express the latter it is also useful to define the following combination of the electric field strengths
\begin{align}
   {{\cal G}_{\mua\mub}}^{\Ma+} &\equiv {{\cal H}_{\mua\mub}}^{\Ma+} \; , \nonumber \\
   {{\cal G}_{\mua\mub}}^{\Ma-} &\equiv 
          e^{-1} \, \eta^{\Ma\Mb} \, \epsilon_{\mua\mub\muc\mud} \, 
	    \frac{ \partial {\cal L}_{\text{kin}} } { \partial {\cal H}_{\muc\mud}^{\Mb+} }
	  \nonumber \\ &
          \; = \; - \ft 1 2 \,  \epsilon_{\mua\mub\muc\mud} \, \Im(\tau) M^{\Ma\Mb} \eta_{\Mb\Mc} {\cal H}^{\Mc+\,\muc\mud}
	     - \Re(\tau) {\cal H}^{\Ma+} _{\mua\mub} \; .
   \label{DefG}	     
\end{align}
The importance of ${{\cal G}_{\mua\mub}}^{\Ma-}$ becomes clear in the ungauged theory
obtained from \eqref{LagBosD4} in the limit $g \rightarrow 0$. In this limit the topological term and the potential
vanish and ${{\cal H}_{\mua\mub}}^{\Ma+}$ and ${{\cal H}_{\mua\mub}}^{\Ma-}$ reduce to Abelian field strengths.
Since the magnetic vectors and the two-form gauge fields only appear projected with some combination of
$f_{\aa\Ma\Mb\Mc}$ and $\xi_{\aa\Ma}$ they completely decouple from the Lagrangian at $g=0$.
The equations of motion for the electric vector fields then take the form $\partial_{[\mua} {{\cal G}_{\mub\muc]}}^{\Ma-} = 0$.
In the ungauged theory magnetic vector fields are introduced by hand via ${\cal H}^{\Ma-}_{\mua\mub} = {\cal G}^{\Ma-}_{\mua\mub}$
and ${\cal G}^{\Ma\aa}=({\cal G}^{\Ma+},{\cal G}^{\Ma-})$ and ${\cal H}^{\Ma\aa}$ are
on-shell identical.

Turning back to the gauged theory one finds for general variations of the vector and two-form gauge fields 
that the Lagrangian varies as \cite{deWit:2005ub}
\begin{align}
   e^{-1} \delta {\cal L}&{}_{\text{bos}} = \ft 1 8 g \left( \Theta_{-\Ma\Mb\Mc} \Delta B^{\Mb\Mc}_{\mua\mub}
                                    + \xi_{-\Ma} \Delta B_{\mua\mub}^{+-} + \xi_{+\Ma} \Delta B_{\mua\mub}^{++} \right)
		 \epsilon^{\mua\mub\muc\mud}  \left( {\cal H}_{\muc\mud}^{\Ma-} - {\cal G}_{\muc\mud}^{\Ma-} \right)
		     \nonumber \\ & 
		   + \ft 1 2 (\delta A_\mua^{\Ma+}) \left( g \, \xi_{\ab\Ma} M_{+\ac} D^\mua M^{\ab\ac}
		                + \frac g 2 \, {\Theta_{+\Ma\Mc}}^{\Mb} M_{\Mb\Md} D^\mua M^{\Md\Mc} 
				  - \epsilon^{\mua\mub\muc\mud} \eta_{\Ma\Mb} \, D_\mub {\cal G}^{\Mb-}_{\muc\mud}
				  \right)
		     \nonumber \\ & 
		   + \ft 1 2 (\delta A_\mua^{\Ma-}) \left( g \, \xi_{\ab\Ma} M_{-\ac} D^\mua M^{\ab\ac}
                          + \frac g 2 \, {\Theta_{-\Ma\Mc}}^{\Mb} M_{\Mb\Md} D^\mua M^{\Md\Mc} 
			  + \epsilon^{\mua\mub\muc\mud} \eta_{\Ma\Mb} \, D_\mub {\cal G}^{\Mb+}_{\muc\mud}  \right)
          \nonumber \\ & + \, \text{total derivatives,}
   \label{varyL}			  
\end{align}
where we used the ``covariant variations''
\begin{align}
    \Delta B^{\Ma\Mb}_{\mua\mub} &= \delta B^{\Ma\Mb}_{\mua\mub} 
                       - 2 \epsilon_{\aa\ab} A^{\aa[\Ma}_{[\mua} \, \delta A^{\Mb]\ab}_{\mub]} \; ,
    \nonumber \\ 		       
    \Delta B^{\aa\ab}_{\mua\mub} &= \delta B^{\aa\ab}_{\mua\mub}
                       + 2 \eta_{\Ma\Mb} A^{\Ma(\aa}_{[\mua} \, \delta A^{\ab)\Mb}_{\mub]} \; .
    \label{CovDB}		       
\end{align}
Equation \eqref{varyL} encodes the gauge field equations of motion of the theory.
Variation of the two-form gauge fields
yields a projected version of the duality equation ${\cal H}^{\Ma-}_{\mua\mub} = {\cal G}^{\Ma-}_{\mua\mub}$
between electric and magnetic vector fields. 
From varying the electric vector fields one obtains a field equation for the electric vectors themselves which contains scalar
currents as source terms. Finally, the variation of the magnetic vectors gives a duality equation between scalars and
two-form gauge fields. To make this transparent one needs the modified Bianchi identity for ${\cal H}^{\Ma+}_{\mua\mub}$
which reads
\begin{align}
   D^{\phantom{\Ma}}_{[\mua} {\cal H}^{\Ma+}_{\mub\muc]} 
      &= \frac g 6 \left( {\Theta_-}{}^{\Ma}{}_{\Mc\Md} {\cal H}^{(3)\Mc\Md}_{\mua\mub\muc}
       +{\xi_+}^\Ma {\cal H}_{\mua\mub\muc}^{(3)++} + {\xi_{-}}^\Ma {\cal H}_{\mua\mub\muc}^{(3)+-} \right) \; ,
\end{align}
where the two-form field strengths are given by
\begin{align}
    {\cal H}^{(3)\Ma\Mb}_{\mua\mub\muc} &= 3 \, \partial^{\phantom{\nu}}_{[\mua} B^{\Ma\Mb}_{\mub\muc]} 
                 + 6 \, \epsilon_{\aa\ab} \, A^{\aa[\Ma}_{[\mua} \, \partial^{\phantom{\nu}}_{\mub} A^{\Mb]\ab}_{\muc]} + {\cal O}(g)\; ,
    \nonumber \\ 		       
    {\cal H}^{(3)\aa\ab}_{\mua\mub\muc} &= 3 \, \partial^{\phantom{\nu}}_{[\mua} B^{\aa\ab}_{\mub\muc]}
                  + 6 \, \eta_{\Ma\Mb} \, A^{\Ma(\aa}_{[\mua} \, \partial^{\phantom{\nu}}_\mub A^{\ab)\Mb}_{\muc]} + {\cal O}(g) \; ,
\end{align}
up to terms of order $g$.

Thus we find that the tensors $f_{\aa\Ma\Mb\Mc}$ and $\xi_{\aa\Ma}$ do not only specify the gauge group but also
organize the couplings of the various fields. They determine which vector gauge fields appear in the covariant derivatives,
how the field strengths have to be modified, which magnetic vector fields and which two-form gauge fields enter the Lagrangian
and how they become dual to electric vector fields and scalars via their equation of motion.
However, consistency of the entire construction
above crucially depends on some particular quadratic constraints that $f_{\aa\Ma\Mb\Mc}$
and $\xi_{\aa\Ma}$ have to satisfy and which are presented
in the next subsection.

In principle one should also give the fermionic contributions to the Lagrangian and check supersymmetry to verify
that \eqref{LagBosD4} really describes the bosonic part of a supergravity theory.
We have obtained the results by applying the general method of covariantly coupling electric and magnetic vector gauge fields in 
a gauged theory \cite{deWit:2005ub} to the particular case of $N=4$ supergravity.
This fixes the bosonic Lagrangian up to the scalar potential. The latter is also strongly restricted by gauge invariance,
only those terms that appear in \eqref{VD4} are allowed. We obtained the pre-factors between the various terms
by matching the scalar potential with the one known from half-maximal supergravity in three spacetime dimensions \cite{deWit:2003ja},
see appendix \ref{app:D3}. The general theory then was compared with various special cases that were already
worked out elsewhere \cite{deRoo:1985jh,Bergshoeff:1985ms,
deRoo:2003rm,D'Auria:2002tc,D'Auria:2003jk,Angelantonj:2003rq,
Angelantonj:2003up,Wagemans:1990mv,Villadoro:2004ci,Kaloper:1999yr}, see section \ref{sec:D4examples}.

\subsection{Quadratic constraints and gauge invariance}
\label{sec:D4con}

We have seen that the tensors $\xi_{\aa\Ma}$ and $f_{\aa\Ma\Mb\Mc}=f_{\aa[\Ma\Mb\Mc]}$
parameterize the possible gaugings of the theory.
These are constant tensors (their entries are fixed real numbers) for which we demand
in addition the following set of consistency constraints
\begin{align}
   \xi_\aa^\Ma \xi_{\ab\Ma} &= 0 \, , \nonumber \\[1ex]
   \xi^\Mc_{(\aa}  f_{\ab)\Mc\Ma\Mb} &= 0 \, , \nonumber \\[1ex]
   3 f_{\aa\Me[\Ma\Mb} {f_{\ab\Mc\Md]}}^\Me + 2 \xi_{(\aa[\Ma} f_{\ab)\Mb\Mc\Md]} &= 0 \; , \nonumber \\[1ex]
   \epsilon^{\aa\ab} \left( \xi_{\aa}^\Mc f_{\ab\Mc\Ma\Mb} + \xi_{\aa\Ma} \xi_{\ab\Mb} \right) &= 0 \, , \nonumber \\
   \epsilon^{\aa\ab} \left( f_{\aa\Ma\Mb\Me} {f_{\ab\Mc\Md}}^\Me - \xi^\Me_\aa f_{\ab\Me[\Ma[\Mc} \eta_{\Md]\Mb]}
       - \xi_{\aa[\Ma} f_{\Mb][\Mc\Md]\ab} + \xi_{\aa[\Mc} f_{\Md][\Ma\Mb]\ab} \right) &= 0 \, .
   \label{QConD4}
\end{align}
These quadratic constraints guarantee the closure of the gauge group, as will be explained below.
The deformation of the theory is consistent if and only if these constraints are satisfied.
They are invariant under the global symmetry group:
given one solution one can create another one by a $G$ action. But all solutions generated in
this way describe the same gauged supergravity. This is obvious for those $G$ transformation
that belong to the ${\rm SO}(1,1) \times {\rm SO}(6,n)$ off-shell symmetry since the entire construction
of the last section was formally invariant under these transformations, i.e. these transformations
correspond to a linear field redefinition that does not mix magnetic and electric vector fields.
In contrast, two solutions of the constraints which are related by a general ${\rm SL}(2)$ transformation
yield two theories which at first sight look rather different but are related by a symplectic transformation
which rotates electric into magnetic vector fields and vice versa. 

It is convenient to define a composite index for the vector fields by
$A_\mua{}^\sMa = A_\mua{}^{\Ma\aa}$, and a symplectic form $\Omega_{\sMa\sMb}$ by
\begin{align}
   \Omega_{\sMa\sMb} \, &= \, \Omega_{\Ma\aa \, \Mb\ab} \, \equiv \, \eta_{\Ma\Mb} \epsilon_{\aa\ab} \, , &
   \Omega^{\sMa\sMb} \, &= \, \Omega^{\Ma\aa \, \Mb\ab} \, \equiv \, \eta^{\Ma\Mb} \epsilon^{\aa\ab} \, ,
\end{align}
The symplectic group ${\rm Sp}(12+2n)$ is the group of linear transformations that preserve $\Omega_{\sMa\sMb}$.
An arbitrary symplectic rotation of the theory gives a Lagrangian that is not yet contained in the description above
but which describes the same theory on the level of the equations of motion. All possible
Lagrangians of gauged $N=4$ supergravity are thus parameterized by $\xi_{\aa\Ma}$, $f_{\aa\Ma\Mb\Mc}$ and
an element of ${\rm Sp}(12+2n)$.

In order to illustrate the meaning of the quadratic constraints \eqref{QConD4} we first consider the case
of purely electric gaugings for which $\xi_{\aa\Ma}=0$ and $f_{-\Ma\Mb\Mc}=0$.
In this case only electric vector fields $A_\mua{}^{\Ma+}$
enter the Lagrangian. We then find
${f_{+\Ma\Mb}}^\Mc = f_{+\Ma\Mb\Md} \, \eta^{\Md\Mc}$ to be the structure constants of
the gauge group and the constraint \eqref{QConD4} simplifies to the Jacobi identity
\begin{align}
   f_{+\Me[\Ma\Mb} {f_{+\Mc\Md]}}^\Me &= 0 \, .
   \label{JacobiFP}
\end{align}
Due to this identity the topological term ${\cal L}_{\text{top}}$ vanishes in this case.
Note that the ${\rm SO}(6,n)$ metric $\eta_{\Ma\Mb}$ is used to contract the indices in \eqref{JacobiFP},
while in the ordinary Jacobi identity the Cartan Killing form occurs.  Also the indices
$\Ma,\Mb,\ldots$ run over $6+n$ values while the gauge group might be of smaller dimension.
These issues will be discussed in section \ref{sec:D4examples}.

In the general case of an arbitrary solution of \eqref{QConD4} we can read off the gauge group generators
from the covariant derivative \eqref{CovDivD4}. For an object in the vector field representation we want
\begin{align}
   D_\mua \, \Lambda^\sMa &= \nabla_\mua \, \Lambda^\sMa + g \, A_\mua^\sMb \, X_{\sMb\sMc}{}^\sMa  \, \Lambda^\sMc \; ,
\end{align}
which yields
\begin{align}
   {X_{\sMa\sMb}}^\sMc &= {X_{\Ma\aa \, \Mb \ab}}^{\Mc\ac}  
   \nonumber \\ &= -  \delta_\ab^\ac \, {f_{\aa\Ma\Mb}}^\Mc
      + \frac 1 2 \left( \delta_\Ma^\Mc \, \delta_\ab^\ac \, \xi_{\aa\Mb} -  \delta_\Mb^\Mc \, \delta_\aa^\ac \, \xi_{\ab\Ma} 
      - \delta_\ab^\ac \, \eta_{\Ma\Mb} \, \xi_{\aa}^\Mc
      + \epsilon_{\aa\ab} \, \delta_\Mb^\Mc \, \xi_{\ad\Ma} \, \epsilon^{\ad\ac} \right) \; .
  \label{DefXD4}      
\end{align}
Note that these objects satisfy
\begin{align}
   {X_{\sMa[\sMb}}^\sMd \Omega_{\sMc]\sMd} &= 0  \; , &
   {X_{(\sMa\sMb}}^\sMd \Omega_{\sMc)\sMd} &= 0  \; .
\end{align}
It was found in \cite{deWit:2005ub} that the last of these equations is crucial for consistency of the gauged theory.
It is this linear constraint that demands the gauge group generators to be parameterized by $f_{\aa\Ma\Mb\Mc}$
and $\xi_{\aa\Ma}$ according to \eqref{DefXD4}.

An infinitesimal gauge transformation is parameterized by $\Lambda^{\sMa}(x)=\Lambda^{\Ma\aa}(x)$
and acts on objects $x^\sMa$ and $x_{\sMa}$ in the (dual) vector field representations as
\begin{align}
   \delta x^\sMa &= - g \, \Lambda^\sMb \, {X_{\sMb\sMc}}^\sMa \, x^\sMc \; , &
   \delta x_\sMa &= g \, \Lambda^\sMb \, {X_{\sMb\sMa}}^\sMc \, x_\sMc \; ,
\end{align}
where $g$ is the gauge coupling constant.
This defines the gauge group $G_0 \subset G \subset {\rm Sp}(12+2n)$. Treating the generators
${X_{\sMa\sMb}}^\sMc = {(X_{\sMa})_\sMb}^\sMc$ as matrices we find the following commutator relations
to be satisfied
\begin{align}
   [ X_\sMa , X_\sMb ] &= - {X_{\sMa\sMb}}^\sMc \, X_\sMc \; ,
   \label{ClosureD4}
\end{align}
i.e. the gauge group $G_0$ is closed. Some computation
reveals that the last equation is equivalent to the quadratic constraint \eqref{QConD4}. Therefore the quadratic constraint is
a generalization of the Jacobi identity \eqref{JacobiFP} guaranteeing the closure of the gauge group.
Furthermore according to \eqref{ClosureD4} the generators ${X_{\sMa\sMb}}^\sMc$ take the role of generalized
structure constants. However, they are only antisymmetric in $\sMa$, $\sMb$ after
having contracted with $X_\sMc$. The fact that ${X_{(\sMa\sMb)}}^\sMc$ is in general not vanishing 
explains the need for the two-form gauge fields in the generalized field strengths \eqref{FieldStrD4}.
The ordinary field strength would not transform covariantly under gauge transformations
$\Lambda^{\Ma\aa}(x)$.

The two-form gauge fields $B_{\mua\mub}^{\Ma\Mb}$ and $B_{\mua\mub}^{\aa\ab}$
are equipped with tensor gauge transformations parameterized by
$\Xi^{\Ma\Mb}_\mua=\Xi^{[\Ma\Mb]}_\mua$ and $\Xi^{\aa\ab}_\mua=\Xi^{(\aa\ab)}_\mua$.
Under general vector and tensor gauge transformations the gauge fields transform as
\begin{align}
   \delta A_{\mua}^{\Ma+} &= D_\mua \Lambda^{\Ma+} 
	    - \frac g 2 \, {{\Theta_{-}}^{\Ma}}_{\Mb\Mc} \Xi_{\mua}^{\Mb\Mc}
	    - \frac g 2 \, {\xi_+}^\Ma \Xi_{\mua}^{++} - \frac g 2 {\xi_{-}}^\Ma \Xi_{\mua}^{+-} \; ,
   \nonumber \\	    
   \delta A_{\mua}^{\Ma-} &= D_\mua \Lambda^{\Ma-} 
	    + \frac g 2 \, {{\Theta_{+}}^{\Ma}}_{\Mb\Mc} \Xi_{\mua}^{\Mb\Mc}
	    - \frac g 2 \, {\xi_-}^\Ma \Xi_{\mua}^{--} - \frac g 2 {\xi_{+}}^\Ma \Xi_{\mua}^{+-} \; ,
   \nonumber \\	    
    \Delta B^{\Ma\Mb}_{\mua\mub} &= 2 D_{[\mua} \Xi^{\Ma\Mb}_{\mub]} 
                       - 2 \epsilon_{\aa\ab} \Lambda^{\aa[\Ma} \, {\cal G}^{\Mb]\ab}_{\mua\mub} \; ,
    \nonumber \\ 		       
    \Delta B^{\aa\ab}_{\mua\mub} &= 2 D_{[\mua} \Xi^{\aa\ab}_{\mub]}
                       + 2 \eta_{\Ma\Mb} \Lambda^{\Ma(\aa} \, {\cal G}^{\ab)\Mb}_{\mua\mub} \; ,
\end{align}
where we used the covariant variations of the two-form gauge fields \eqref{CovDB}. Under these gauge transformations
the Lagrangian \eqref{LagBosD4} is invariant.
The only non-vanishing commutator of these gauge transformations is\footnote{
In the Lagrangian the two-form gauge fields only appear under a particular projection with $f_{\aa\Ma\Mb\Mc}$
and $\xi_{\aa\Ma}$ and the gauge transformation on them only close under this very projection
\cite{deWit:2004nw}.}
\begin{align}
   [ \delta_{\Lambda_1} , \delta_{\Lambda_2} ] &= \delta_{\tilde \Lambda} + \delta_{\tilde \Xi} \; ,
   \label{D4gaugealgebra}   
\end{align}
where
\begin{align}
   \tilde \Lambda^\sMa &= g {X_{\sMb\sMc}}^\sMa \Lambda^\sMb_{[1} \Lambda^\sMc_{2]} \, , \nonumber \\
   \tilde \Xi_\mua^{\Ma\Mb} &= \epsilon_{\aa\ab} \left( \Lambda_1^{\aa[\Ma} \, D_\mua \Lambda_2^{\Mb]\ab}
                                                       - \Lambda_2^{\aa[\Ma} \, D_\mua \Lambda_1^{\Mb]\ab} \right)
    \, , \nonumber \\						       
   \tilde \Xi_\mua^{\aa\ab} &= - \eta_{\Ma\Mb} \left( \Lambda_1^{\Ma(\aa} \, D_\mua \Lambda_2^{\ab)\Mb}
                                                     - \Lambda_2^{\Ma(\aa} \, D_\mua \Lambda_1^{\ab)\Mb}  \right) \; .
   \label{GAlgebraD4}						     
\end{align}
In the action on objects that do not transform under tensor gauge transformations (like field strengths, scalar fields) 
this algebra coincides with \eqref{ClosureD4}.

\subsection{Killing spinor equations}
\label{sec:D4kill}

So far we have only considered bosonic fields and we do not intend to give the entire fermionic Lagrangian
nor the complete supersymmetry action. They can e.g. be found in the paper of Bergshoeff, Koh and Sezgin \cite{Bergshoeff:1985ms}
for purely electric gaugings when only $f_{+\Ma\Mb\Mc}$ is non-zero, and we have chosen most of our conventions
to agree with their work in this special case\footnote{
The structure constants $f_{\Ma\Mb\Mc}$ in \cite{Bergshoeff:1985ms} equal minus $f_{+\Ma\Mb\Mc}$.}.
In particular all terms of order $g^0$, 
i.e. terms of the ungauged theory, can be found there.

Our aim in this section is to give the
Killing spinor equations of the general gauged theory,
i.e. the variations of the gravitini and of the spin $1/2$ fermions under supersymmetry.
Those are required for example when studying BPS solutions or when analyzing the supersymmetry breaking or preserving
of particular ground states.

All the fermions carry a representation of  $H={\rm SO}(2) \times {\rm SO}(6) \times {\rm SO}(n)$
which is the maximal compact subgroup of $G$.
Instead of ${\rm SO}(6)$ we work with its covering group ${\rm SU}(4)$ in the following.
The gravity multiplet contains four gravitini $\psi_\mua^\ja$ and four spin $1/2$ fermions $\chi^\ja$ and
in the $n$ vector multiplet there are $4n$ spin 1/2 fermions $\lambda^{\xa\ja}$, where
$\ja=1,\ldots, 4$ and $\xa=1,\ldots,n$ are vector indices of ${\rm SU}(4)$ and ${\rm SO}(n)$.
The ${\rm SO}(2)={\rm U}(1)$ acts on the fermions as a multiplication with a complex phase
$\exp(i q \lambda(x))$, where the charges $q$ are given in table \ref{FermRepD4}.

\begin{table}[tb]
   \begin{center}
     \begin{tabular}{r|ccc}
            & ${\rm SO}(2)$ charges & ${\rm SU}(4)$ rep. & ${\rm SO}(n)$ rep. 
\\ \hline
        gravitini $\psi_\mua^\ja$ & $- \, \ft 1 2$ & ${\bf 4}$ & ${\bf 1}$ \\
	spin $1/2$ fermions $\chi^\ja$~ & $+ \, \ft 3 2$ & ${\bf 4}$ & ${\bf 1}$ \\
	spin $1/2$ fermions $\lambda^{\xa\ja}$ & $+ \, \ft 1 2$ & ${\bf 4}$ & ${\bf n}$
     \end{tabular}
     \caption{\label{FermRepD4}{ \small $H$-representations of the fermions.}}
   \end{center}     
\end{table}

As usual we use gamma-matrices with
\begin{align}
   \{ \Gamma_\mua, \Gamma_\mub \} &= 2 \eta_{\mua\mub} \; , &
   (\Gamma_\mua)^\dag &= \eta^{\mua\mub} \Gamma_{\mub} \; , &
   \Gamma_5 &= i \Gamma_0 \Gamma_1 \Gamma_2 \Gamma_3 \; .
\end{align}
All our fermions are chiral. We choose $\psi_\mua^\ja$ and $\lambda^{\xa\ja}$ to be right-handed while $\chi^\ja$ is left-handed,
that is
\begin{align}
   \Gamma_5 \psi_\mua^\ja &= + \psi_\mua^\ja \; , &
   \Gamma_5 \chi^\ja &= - \chi^\ja \; , &
   \Gamma_5 \lambda^{\xa\ja} &= + \lambda^{\xa\ja} \; . &
\end{align}
Vector indices of ${\rm SU}(4)$ are raised and lowered by complex conjugation, i.e. for an ordinary ${\rm SU}(4)$ vector
$v_\ja = (v^\ja)^*$.
However, for fermions we need the matrix $B=i \Gamma_5 \Gamma_2$
to define $\phi_\ja = B (\phi^\ja)^*$. This ensures
that $\phi_\ja$ transforms as a Dirac spinor when $\phi^\ja$ does. The complex conjugate of a chiral spinor has
opposite chirality, e.g. $\chi_\ja=B (\chi^\ja)^*$ is right-handed\footnote{
Right-handed spinors can be described by Weyl-spinors $\phi^A$, and left-handed ones then turn to conjugate Weyl-spinors
$\phi_{\dot A}$. Here $A$ and $\dot A$ are (conjugate) ${\rm SL}(2,\mathbb{C})$ vector indices.
In the chiral representation of the Gamma-matrices
\begin{align*}
  \Gamma_\mu&=\left(\begin{array}{ccc} 0 & \sigma^\mu \\
          \sigma_{\mu} & 0 \end{array}\right) \; , &
  \Gamma_5&=\left(\begin{array}{ccc} \mathbbm{1} & 0 \\
           0 & - \mathbbm{1} \end{array}\right) \; , &
   B &= i \Gamma_5 \Gamma_2 =  \left(\begin{array}{ccc} 0 & \epsilon \\
          - \epsilon & 0 \end{array}\right) \; ,
\end{align*}
where $\epsilon$ is the two-dimensional epsilon-tensor and $\sigma_\mua=(\mathbbm{1},\vec \sigma)$,
$\sigma^\mua=\eta^{\mua\mub} \sigma_{\mub}=(-\mathbbm{1},\vec \sigma)$ contains the Pauli matrices,
we find right-handed spinors to have the form $\phi=(\phi^A,0)^T$ while left-handed ones look like $\phi=(0,\phi_{\dot A})^T$.
Thus we have $\chi^\ja=(0,\chi^\ja_{\dot A})^T$ and its complex conjugate is given by $\chi_\ja=(\chi_\ja^A,0)^T$ where
the Weyl-spinors are related by $\chi_\ja^A=\epsilon^{AB} (\chi^\ja_{\dot B})^*$.}.
For $\bar \phi{}_\ja = (\phi^\ja)^\dag \Gamma_0$ we define the complex conjugate by
$\bar \phi{}^\ja = (\bar \phi{}_\ja)^* B$ which yields
$\bar \phi{}_\ja \chi^\ja = \bar \chi{}^\ja \phi_\ja = (\bar \phi{}^\ja \chi_\ja)^* = (\bar \chi{}_\ja \phi^\ja)^*$.

An ${\rm SO}(6)$ vector $v^\ya$ can alternatively be described by an antisymmetric tensor
$v^{\ja\jb}=v^{[\ja\jb]}$ subject to the pseudo-reality constraint
\begin{align}
   v_{\ja\jb} &= (v^{\ja\jb})^* = \frac 1 2 \epsilon_{\ja\jb\jc\jd} v^{\jc\jd} \; .
\end{align}
We normalize the map $v^\ya \mapsto v^{\ja\jb}$ such that the scalar product becomes
\begin{align}
   v^\ya w^\ya &= \frac 1 2 \epsilon_{\ja\jb\jc\jd}  v^{\ja\jb} w^{\jc\jd} \; .
\end{align}   
We can thus rewrite the coset representative ${{\cal V}_\Ma}^{\ya}$ as ${{\cal V}_\Ma}^{\ja\jb}$
such that the equations \eqref{DefEta} and \eqref{DefM6} become
\begin{align}
   \eta_{\Ma\Mb} &= - \frac 1 2 \epsilon_{\ja\jb\jc\jd} {{\cal V}_\Ma}^{\ja\jb} {{\cal V}_\Mb}^{\jc\jd} 
                     + {{\cal V}_\Ma}^\xa {{\cal V}_\Mb}^\xa \; ,
   \nonumber \\
   M_{\Ma\Mb\Mc\Md\Me\Mf}  
         &= -  \, 2 \, i \, \epsilon_{\ja\jb\jg\jj} \, \epsilon_{\jc\jd\jh\jk} \, \epsilon_{\je\jf\ji\jl} \,
	         {{\cal V}_{[\Ma}}^{\ja\jb} {{\cal V}_\Mb}^{\jc\jd} {{\cal V}_\Mc}^{\je\jf} 
                             {{\cal V}_\Md}^{\jg\jh} {{\cal V}_\Me}^{\ji\jj} {{\cal V}_{\Mf]}}^{\jk\jl}  \; .
\end{align}
The scalar matrices ${{\cal V}_\Ma}^{\ja\jb}$ and ${{\cal V}_\Ma}^{\xa}$ can be used to translate from ${\rm SO}(6,n)$
representations
under which the vector and tensor gauge fields transform into ${\rm SO}(6) \times {\rm SO}(n)$ representations carried by the
fermions. They are thus crucial when we want to couple fermions. For the same reason it is necessary to introduce
an ${\rm SL}(2)$ coset representative, namely a complex ${\rm SL}(2)$ vector ${\cal V}_{\aa}$ which satisfies
\begin{align}
   M_{\aa\ab} &= \Re( {\cal V}_{\aa} ({\cal V}_{\ab})^* ) \; .
\end{align}
Under ${\rm SO}(2)$ ${\cal V}_{\aa}$ carries charge $+1$ while its complex conjugate carries charge $-1$.\footnote{
The complex scalars $\phi$ and $\psi$ in \cite{Bergshoeff:1985ms} translate into our notation as
${\cal V}_+ = \psi$, ${\cal V}_- = i \phi$ and $\psi/\phi=i \tau^*$.}

When gauging the general theory all partial derivatives are replaced by covariant derivatives $\partial \rightarrow D$ and
all Abelian field strengths by covariant ones ${\cal F}^{\Ma+} \rightarrow {\cal H}^{\Ma+}$. Moreover one has to add the
topological term and the scalar potential to the Lagrangian as we have described in section \ref{sec:D4lag}.
In the fermionic sector the only additional change that has to be made in the Lagrangian is the introduction of fermionic
mass terms and fermionic couplings, all of order $g^1$. For example those terms that involve the gravitini read
\begin{align}
   e^{-1} {\cal L}_{\text{f.mass}} \, &=
                           \, \ft 1 3 \, g \, A_1^{\ja\jb} \, \bar \psi{}_{\mua\ja} \, \Gamma^{\mua\mub} \, \psi_{\mub\jb}
                               - \ft 1 3 \, i \, g \, A_2^{\ja\jb} \, \bar \psi{}_{\mua\ja} \, \Gamma^{\mua} \, \chi_{\jb}
			       + i g \, {A_{2\,\xa\ja}}^\jb \, \bar \psi{}^\ja_{\mua}  \, \Gamma^{\mua} \, \lambda^\xa_{\jb}
			       + \text{h.c.} \; ,
\end{align}
where $A_1^{\ja\jb}=A_1^{(\ja\jb)}$, $A_2^{\ja\jb}$ and ${A_{2\,\xa\ja}}^\jb$ are the so called fermion shift matrices
which depend on the scalar fields.

Also the supersymmetry transformations of the fermions have to be endowed with corrections of order $g^1$, namely
\begin{align}
   \delta \psi_\mua^\ja &= 2 D_\mua \epsilon^\ja
             + \ft 1 4 \, i \, ({\cal V}_\aa)^* {{\cal V}_\Ma}^{\ja\jb} \, {\cal G}^{\Ma\aa}_{\mub\muc}  
	                              \Gamma^{\mub\muc} \Gamma_\mua \epsilon_\jb
	      - \ft 2 3 \, g \, A_1^{\ja\jb} \Gamma_\mua \epsilon_\jb \, , \nonumber \\
   \delta \chi^\ja &= i \, \epsilon^{\aa\ab} {\cal V}_\aa (D_\mua {\cal V}_\ab) \Gamma^\mua \epsilon^\ja
                + \ft 1 2 \, i \, {\cal V}_\aa {{\cal V}_\Ma}^{\ja\jb} \, {\cal G}^{\Ma\aa}_{\mua\mub} \Gamma^{\mua\mub} \epsilon_\jb 
		     - \ft 4 3 \, i \, g  \, A_2^{\jb\ja} \epsilon_\jb \, , \nonumber \\
   \delta \lambda_\xa^{\ja} &= 2 i \, {{\cal V}_{\xa}}^\Ma ( D_\mua {{\cal V}_\Ma}^{\ja\jb} ) \Gamma^\mua \epsilon_{\jb} 
		    - \ft 1 4 \, {\cal V}_\aa {\cal V}_{\Ma\xa} \, {\cal G}^{\Ma\aa}_{\mua\mub} \Gamma^{\mua\mub} \epsilon^\ja
                               + 2 \, i \, g \, A_{2\,\xa\jb}{}^{\ja}  \, \epsilon^\jb    \; ,
   \label{varyF}			       
\end{align}
where the same matrices $A_1$ and $A_2$ appear as in the Lagrangian. 
There are also higher order fermion terms in the supersymmetry rules, but those do not get corrections in the gauged theory.
We wrote the vector field contribution to the
fermion variations in an ${\rm SL}(2)$ covariant way. Using the definition \eqref{DefG} one finds
\begin{align}
   i \, {\cal V}_\aa {{\cal V}_\Ma}^{\ja\jb} {\cal G}^{\Ma\aa}_{\mua\mub} \Gamma^{\mua\mub}
      &= ({\cal V}_-{}^*)^{-1} \, {{\cal V}_\Ma}^{\ja\jb} \left( {\cal H}^{\Ma+}_{\mua\mub} 
                      + \ft 1 2 \, i \, \epsilon_{\mua\mub\muc\mud} {\cal H}^{\Ma+\,\muc\mud} \right) \Gamma^{\mua\mub}
     \nonumber \\ &		      
       = ({\cal V}_-{}^*)^{-1} \, {{\cal V}_\Ma}^{\ja\jb}  {\cal H}^{\Ma+}_{\mua\mub} \Gamma^{\mua\mub} ( 1 - \Gamma_5 )  
       \, , \nonumber \\
   i \, {\cal V}_\aa {{\cal V}_\Ma}^{\xa} {\cal G}^{\Ma\aa}_{\mua\mub} \Gamma^{\mua\mub}
      &= ({\cal V}_-{}^*)^{-1} \, {{\cal V}_\Ma}^{\xa} \left( {\cal H}^{\Ma+}_{\mua\mub} 
                      - \ft 1 2 \, i \, \epsilon_{\mua\mub\muc\mud} {\cal H}^{\Ma+\,\muc\mud} \right) \Gamma^{\mua\mub}
     \nonumber \\ &		      
       = ({\cal V}_-{}^*)^{-1} \, {{\cal V}_\Ma}^{\xa}  {\cal H}^{\Ma+}_{\mua\mub} \Gamma^{\mua\mub} ( 1 + \Gamma_5 )  \; .
\end{align}
Explicitly, the fermion shift matrices are given by
\begin{align}
   A_1^{\ja\jb} &= \epsilon^{\aa\ab} ({\cal V}_\aa)^* 
                   {{\cal V}_{[\jc\jd]}}^\Ma {{\cal V}_\Mb}^{[\ja\jc]} {{\cal V}_\Mc}^{[\jb\jd]}  {f_{\ab\Ma}}^{\Mb\Mc}  \; ,
		   \nonumber \\
   A_2^{\ja\jb} &= \epsilon^{\aa\ab} {\cal V}_\aa
                    {{\cal V}_{[\jc\jd]}}^\Ma {{\cal V}_\Mb}^{[\ja\jc]} {{\cal V}_\Mc}^{[\jb\jd]}  {f_{\ab\Ma}}^{\Mb\Mc} 
		   + \frac 3 2 \epsilon^{\aa\ab} {\cal V}_\aa {{\cal V}_\Ma}^{\ja\jb} {\xi_\ab}^\Ma  \; ,
		    \nonumber \\
   {A_{2\, \xa\ja}}^\jb &= \epsilon^{\aa\ab} {\cal V}_\aa
                     {{\cal V}_\Ma}^{\xa} {{\cal V}^\Mb}_{[\ja\jc]} {{\cal V}_\Mc}^{[\jb\jc]} {f_{\ab\Ma\Mb}}^\Mc 
		    - \frac 1 4 \delta_\ja^\jb \epsilon^{\aa\ab} {\cal V}_\aa {{\cal V}_\xa}^{\Ma} \xi_{\ab\Ma}  \; .
\end{align}
Supersymmetry of the Lagrangian forces them to obey in particular\footnote{
This equation is obtained by considering terms of the
form $g^2 \bar \psi_\mua \Gamma^\mua \epsilon$ in the variation $\delta {\cal L}$.}
\begin{align}
    \ft 1 3 \, A_1^{\ja\jc} \, {\bar A}_{1\,\jb\jc}  - \, \ft 1 9 \, A_2^{\ja\jc} \, {\bar A}_{2\,\jb\jc}
               - \, \ft 1 2 \, {A_{2\, \xa\jb}}^\jc \, {\bar A}_{2\, \xa}{}^\ja{}_\jc \,  &= \, - \, \ft 1 4 \, \delta^\ja_\jb \, V \;,
    \label{WardD4}	       
\end{align}
where the scalar potential $V$ appears on the right hand side.
The last equation is indeed satisfied as a consequence of the quadratic constraints \eqref{QConD4}.

If we have chosen $f_{\aa\Ma\Mb\Mc}$ and $\xi_{\aa\Ma}$ such that the scalar potential possesses an extremal point
one may wonder whether the associated ground state conserves some supersymmetry, i.e. whether
$\epsilon^\ja$ exists such the fermion variations \eqref{varyF} vanish in the ground state.
The usual Ansatz is $\epsilon^\ja=q^\ja \, \xi$, where $q^\ja$ is just an ${\rm SU}(4)$ vector
while $\xi$ is a right-handed Killing spinor of AdS ($V<0$) or Minkowski ($V=0$) space, i.e.\footnote{
Consistency of the AdS Killing spinor equation can be checked by using
$R_{\mua\mub\muc\mud}=- \ft 2 3 g^2 V g_{\mua[\muc} g_{\mud]\mub}$, $\Gamma_{[\mua} B \Gamma_{\mub]}^* B^* = - \Gamma_{\mua\mub}$
and $[D_\mua,D_\mub] \xi=-\ft 1 4 {R_{\mua\mub}}^{\muc\mud} \Gamma_{\muc\mud} \xi$.}
\begin{align}
   D_\mua \xi &= g \sqrt{- \ft 1 {12} V} \,  \Gamma_\mua B \xi^* \;.
\end{align}
The Killing spinor equations $\delta \psi^\ja = 0$, $\delta \chi^\ja = 0$ and  $\delta \lambda^{\xa\ja} = 0$ then take the form
\begin{align}
   A_1^{\ja\jb} q_\jb &= \sqrt{- \ft 3 4 V} \, q^\ja \; , &
   q_\jb A_2^{\jb\ja} &= 0 \; , &
   {A_{2\xa\jb}}^\ja q^\jb &= 0 \; .
   \label{KillingD4}
\end{align}
Due to \eqref{WardD4} the first equation of \eqref{KillingD4} already implies the other two.

\subsection{Examples}
\label{sec:D4examples}

In this section we give examples of tensors $f_{\aa\Ma\Mb\Mc}$ and $\xi_{\aa\Ma}$ that solve the constraints \eqref{QConD4},
therewith giving examples of gauged $N=4$ supergravities. We recover
those gaugings that were already discussed in the literature but also obtain new ones.

\subsubsection{Purely electric gaugings}

It can be shown that as a consequence of the constraints \eqref{QConD4} for every consistent gauging 
one can perform a symplectic rotation such that only the electric vector fields
serve as gauge fields \cite{deWit:2005ub}. In the maximal supersymmetric theory,
i.e. for $N=8$, this statement can even be reversed,
i.e. every gauging (defined by some embedding tensor similar to our $f_{\aa\Ma\Mb\Mc}$ and $\xi_{\aa\Ma}$) that is purely
electric in some symplectic frame is consistent (i.e. solves the quadratic constraints for the embedding tensor).
This is different in $N=4$ where a nontrivial quadratic constraint remains also for purely electric gaugings.

In the particular electric frame we have chosen -- the one in which the electric and
magnetic vector fields each form a vector under ${\rm SO}(6,n)$ -- the purely electric gaugings are those
for which $f_{-\Ma\Mb\Mc}=0$ and $\xi_{\aa\Ma}=0$, thus only $f_{+\Ma\Mb\Mc}$ is non-vanishing.
This is the class of theories that were constructed by Bergshoeff, Koh and Sezgin \cite{Bergshoeff:1985ms}.
As mentioned above the quadratic constraint in this case simplifies to
the Jacobi identity \eqref{JacobiFP}, which may alternatively be written as
\begin{align}
   {f_{+\Me[\Ma}}^\Md {f_{+\Mb\Mc]}}^\Me &= 0 \, .
   \label{JacobiFP2}
\end{align}
This is a constraint on ${f_{+\Ma\Mb}}^\Mc=f_{+\Ma\Mb\Md} \eta^{\Md\Mc}$ only, but in addition  
the linear constraint $f_{+\Ma\Mb\Mc}=f_{+[\Ma\Mb\Mc]}$ has to be satisfied, such that 
the ${\rm SO}(6,n)$ metric $\eta_{\Ma\Mb}$ enters
non-trivially into this system of constraints. The dimension of the gauge group can at most be $6+n$, which is obvious in
the case that we consider here ($\Ma=1,\ldots,6+n$), but which is also the general limit for arbitrary gaugings.

We first want to consider semi-simple gaugings. Let ${f_{ab}}^c$ be the structure constants of a semi-simple gauge group $G_0$,
where $a,b,c=1 \ldots \dim(G_0)$, $\dim(G_0)\leq 6+n$, then $\eta_{ab}={f_{ac}}^d {f_{bd}}^c$  is the Cartan-Killing form and
we can choose a basis such that it becomes diagonal, i.e.
\begin{align}
  \eta_{ab} &= \diag(\,\underbrace{1, \dots,}_{p}\underbrace{-1,\dots}_{q} \,) \;.
\end{align}
We can only realize the gauge group $G_0$ if we can embed its Lie algebra $\mathfrak{g}_0=\{v^a\}$ into the vector space 
of electric vector fields such that the preimage of $\eta_{\Ma\Mb}$ agrees with $\eta_{ab}$ up to a factor. This
puts a restriction on the signature of $\eta_{ab}$, namely either $p \leq 6$, $q \leq n$ (case 1) or $p \leq n$, $q \leq 6$
(case 2). To make the embedding explicit we define the index $\hat M$ with range $\hat M = 1 \ldots p , 7 \ldots 6+q$
(case 1) or $\hat M = 1 \ldots q , 7 \ldots 6+p$ (case 2). We then have $(\eta_{\hat M \hat N}) = \pm (\eta_{ab})$ and
we can define
\begin{align}
   ( f_{+\hat M \hat N \hat P} ) &= ( f_{abc} ) \; , \qquad \text{all other entries of $f_{+\Ma\Mb\Mc}$ zero,}
\end{align}
where $f_{abc} = {f_{ab}}^d \eta_{dc}$. Since $G_0$ is semi-simple $f_{abc}$ is completely antisymmetric and thus
$f_{+\Ma\Mb\Mc}$ satisfies the linear and the quadratic constraint. For $n \leq 6$ the possible simple groups
that can appear as factors in $G_0$ are
${\rm SU}(2)$, ${\rm SO}(2,1)$, ${\rm SO}(3,1)$, ${\rm SL}(3)$, ${\rm SU}(2,1)$, ${\rm SO}(4,1)$ and ${\rm SO}(3,2)$.
For larger $n$ we then find ${\rm SU}(3)$, ${\rm SO}(5)$, ${\rm G}_{2(2)}$, ${\rm SL}(4)$, ${\rm SU}(3,1)$,
${\rm SO}(5,1)$, etc.

Apart from these semi-simple gaugings there are various non-semi-simple gaugings that satisfy \eqref{JacobiFP2}.
Of those we only want to give an example. We can choose three mutual orthogonal lightlike vectors $a_\Ma$, $b_\Ma$ and $c_\Ma$
and define $f_{+\Ma\Mb\Mc}$ to be the volume form on their span, i.e.
\begin{align}
   f_{+\Ma\Mb\Mc} &= a_{[\Ma} b_\Mb c_{\Mc]} \; .
\end{align}
The vectors have to be linearly independent in order that $f_{+\Ma\Mb\Mc}$ is non-vanishing.
The quadratic constraint is then satisfied trivially since it contains $\eta_{\Ma\Mb}$ which is vanishing on the domain
of $f_{+\Ma\Mb\Mc}$. The gauge group turns out to be $G_0={\rm U}(1)^3$. We can generalize this construction by choosing
$f_{+\Ma\Mb\Mc}$ to be any three form that has as domain a lightlike subspace of the vector space $\{v^\Ma\}$.
All corresponding gauge groups are Abelian.

None of the purely electric gaugings can have a ground state 
with non-vanishing cosmological constant
since the scalar potential \eqref{VD4} in this case is proportional to
$M^{++}=\Im(\tau)^{-1}$. Therefore de Roo and Wagemans introduced a further deformation of the theory \cite{deRoo:1985jh}.
Starting from a semi-simple gauging as presented
above they introduced a phase for every simple group factor as additional parameters in the description of the gauging.
In the next section we will explain the relation of these phases to our parameters $f_{\aa\Ma\Mb\Mc}$
and show how these theories fit into our framework.

\subsubsection{The phases of de Roo and Wagemans}

We now allow for $f_{+\Ma\Mb\Mc}$ and $f_{-\Ma\Mb\Mc}$ to be non-zero but keep $\xi_\aa^\Ma=0$. The
quadratic constraint \eqref{QConD4} then reads
\begin{align}
   f_{\aa\Me[\Ma\Mb} {f_{\ab\Mc\Md]}}^\Me &= 0 \; ,\qquad\qquad
   \epsilon^{\aa\ab} f_{\aa\Ma\Mb\Me} {f_{\ab\Mc\Md}}^\Me = 0 \, .
   \label{QconRW}
\end{align}
To find solutions we start from the situation of the last section, i.e. we assume to have some structure constants
$f_{\Ma\Mb\Mc}=f_{[\Ma\Mb\Mc]}$ that satisfy the Jacobi-identity ${f_{\Me[\Ma}}^\Md {f_{\Mb\Mc]}}^\Me = 0$. In
addition we assume to have a decomposition of the vector space $\{v^\Ma\}$ into $K$ mutual orthogonal subspaces with projectors
$\mathbbm{P}_{i\Ma}{}^\Mb$, $i=1 \ldots K$, i.e. such that for a general vector $v_\Ma$ we have
\begin{align}
   v_\Ma &= \sum_{i=1}^K \, \mathbbm{P}_{i\Ma}{}^\Mb v_\Mb \; , &
   \eta^{\Ma\Mc} \, \mathbbm{P}_{i\Ma}{}^\Mb \, \mathbbm{P}_{j\Mc}{}^\Md &= 0 \qquad \text{for } i \neq j \; .
\end{align}
Furthermore this decomposition shall be such that the three form $f_{\Ma\Mb\Mc}$ does not mix between the subspaces, i.e.
it decomposes into a sum of three-forms on each subspace
\begin{align}
   f_{\Ma\Mb\Mc} &= \sum_{i=1}^K \, f^{(i)}_{\Ma\Mb\Mc} \; , &
   f^{(i)}_{\Ma\Mb\Mc} &= \mathbbm{P}_{i\Ma}{}^\Md \, \mathbbm{P}_{i\Mb}{}^\Me \, \mathbbm{P}_{i\Mc}{}^\Mf \, f_{\Md\Me\Mf} \; .
\end{align}
This implies that the gauge group splits into $K$ factors $G_0=G^{(1)} \times G^{(2)} \times \ldots \times G^{(K)}$ with
$f^{(i)}_{\Ma\Mb\Mc}$ being the structure constant of the $i$-th factor, each of them satisfying the above Jacobi-identity separately.
Solutions of the constraint \eqref{QconRW} are then given by
\begin{align}
   f_{\aa\Ma\Mb\Mc} &= \sum_{i=1}^K \, w^{(i)}_\aa  \, f^{(i)}_{\Ma\Mb\Mc} \; , &
   w^{(i)}_\aa &= ( w^{(i)}_+ , \, w^{(i)}_- ) = ( \cos \alpha_i ,  \, \sin \alpha_i ) ,
   \label{GaugingRW}
\end{align}
where the $w^{(i)}_\aa$ could be arbitrary ${\rm SL}(2)$ vectors which we could restrict to have unit length without
loss of generality. The $\alpha_i \in \mathbbm{R}$, $i=1\ldots K$, are the de Roo-Wagemans-phases
first introduces in \cite{deRoo:1985jh}.
In the following we want to use the abbreviations $c_i = \cos \alpha_i$, $s_i = \sin \alpha_i$.
If $K=1$ we find $f_{+\Ma\Mb\Mc}$ and $f_{-\Ma\Mb\Mc}$ to be proportional.
This case is equivalent to the purely electric gaugings of the last section
since one always finds an ${\rm SL}(2)$ transformation such that $w^{(1)}_\aa$ becomes $(1,0)$.

For a semi-simple gauging as described in the last section there is a natural decomposition of $\{v^\Ma\}$
into mutual orthogonal subspaces and $K$ equals the number of simple factors in $G_0$. But the above construction
also applies for non-semi-simple gaugings.

We have mentioned above that every consistent gauging is purely electric in a particular symplectic frame. 
Considering a concrete gauging it is therefore natural to formulate the theory in this particular frame,
and also the two-form gauge fields then disappear from the Lagrangian.
For those gaugings defined by \eqref{GaugingRW} we may perform the symplectic transformation
\begin{align}
   \tilde A{}^{\Ma+}_\mua &= \sum_{i=1}^K \, c_i \, \mathbbm{P}_i{}^{\Ma}{}_\Mb \, A_\mua^{\Mb+} 
                            +\sum_{i=1}^K \, s_i \, \mathbbm{P}_i{}^{\Ma}{}_\Mb \, A_\mua^{\Mb-} \nonumber \, , \\
   \tilde A{}^{\Ma-}_\mua &= - \sum_{i=1}^K \, s_i \, \mathbbm{P}_i{}^{\Ma}{}_\Mb \, A_\mua^{\Mb+} 
                            +\sum_{i=1}^K \, c_i \, \mathbbm{P}_i{}^{\Ma}{}_\Mb \, A_\mua^{\Mb-} \, ,
\end{align}
such that the covariant derivative depends exclusively on $\tilde A{}_\mua^{\Ma+}$
\begin{align}
    D_\mua      &= \nabla_\mua 
            - g \, \tilde A{}_\mua{}^{\Ma+} {f_{\Ma}}^{\Mb\Mc} t_{\Mb\Mc} \; .
\end{align}
Note that the new electric vector fields $\tilde A{}_\mua^{\Ma+}$ do not form a vector under ${\rm SO}(6,n)$, but transform
into $\tilde A{}_\mua^{\Ma-}$ under this group. The Lagrangian in the new symplectic frame reads
\begin{align}
   e^{-1} {\cal L} &= \, \ft 1 2 \, R 
                    + \, \ft 1 8 \, (D_\mua M_{\Ma\Mb}) (D^\mua M^{\Ma\Mb})
     - \frac 1 {4 \, \Im(\tau)^2}  (D_\mua \tau) (D^\mua \tau^*) 
     \nonumber \\[1ex]      &  \qquad
     - \, \ft 1 4 \, {\cal I}_{\Ma\Mb} {\tilde {\cal F}}{}_{\mua\mub}{}^{\Ma+} \tilde {\cal F}{}^{\mua\mub\Mb+}
               - \, \ft 1 8 \, {\cal R}_{\Ma\Mb}  \, \epsilon^{\mua\mub\muc\mud} 
    {\tilde {\cal F}_{\mua\mub}}^{\Ma+} {\tilde {\cal F}_{\muc\mud}}^{\Mb+}  
    - g^2 V    \;,
\end{align}
and the scalar potential \eqref{VD4} takes the form \cite{Wagemans:1990mv}
\begin{align}
    V  & = \ft 1 {16} \, \Im(\tau)^{-1} \,
      \sum_{i,j=1}^K \left( c_i c_j - 2 \Re(\tau) c_i s_j + |\tau|^2 s_i s_j \right)
      f^{(i)}_{\Ma\Mb\Mc} f^{(j)}_{\Md\Me\Mf} \nonumber \\ & \qquad \qquad \qquad \qquad \qquad \qquad
      \times \Big[\ft 1 3 \, M^{\Ma\Md} M^{\Mb\Me} M^{\Mc\Mf} 
        + ( \ft 2 3 \, \eta^{\Ma\Md} -  M^{\Ma\Md} ) \eta^{\Mb\Me} \eta^{\Mc\Mf}   \Big]
	\nonumber \\ & \qquad 
	        - \ft 1 {18} \, \sum_{i,j=1}^K \, c_i s_j
		             f^{(i)}_{\Ma\Mb\Mc} f^{(j)}_{\Md\Me\Mf} M^{\Ma\Mb\Mc\Md\Me\Mf} \; .
\end{align}
The kinetic term of the vector fields involves the field strength
\begin{align}
   \tilde {\cal F}{}_{\mua\mub}{}^{\Ma+} &= 2 \partial_{[\mua} \tilde A{}_{\mub]}{}^{\Ma+} 
            - g \, f{}_{\Mb\Mc}{}^\Ma \tilde A{}_{[\mua}{}^{\Mb+} \tilde A{}_{\mub]}{}^{\Mc+}  \; ,
\end{align}
and the scalar dependent matrices ${\cal I}_{\Ma\Mb}$ and ${\cal R}_{\Ma\Mb}$ which are defined by
\begin{align}
   ({\cal I}^{-1})^{\Ma\Mb} &= \frac 1 {\Im(\tau)} \sum_{i,j=1}^K 
        \left( c_i c_j - 2 \Re(\tau) c_i s_j + |\tau|^2 s_i s_j \right) 
	  \mathbbm{P}_i{}^{\Ma}{}_\Mc \mathbbm{P}_j{}^{\Mb}{}_\Md \, M^{\Mc\Md} \; , \nonumber \\
  {\cal R}_{\Ma\Mb} ({\cal I}^{-1})^{\Mb\Mc} &= 
    \frac 1 {\Im(\tau)} \sum_{i,j=1}^K 
        \left[ - c_i s_j + \Re(\tau) ( s_i s_j - c_i c_j) + |\tau|^2 s_i c_j \right]
	  \mathbbm{P}_{i\Ma\Mb} \mathbbm{P}_j{}^{\Mc}{}_\Me M^{\Mb\Me} \; .
\end{align}
In general when going to the electric frame for an arbitrary gauging there is still a topological term
for the electric fields of the form $AA\partial A + AAAA$ \cite{deWit:1984px}, but here this term is not present.

Comparing the scalar potential $V$ for non-vanishing phases $\alpha_i$ with that of the last section we find it to have 
a much more complicated $\tau$ dependence and one can indeed find gaugings where it possesses 
stationary points \cite{deRoo:2003rm,Wagemans:1990mv}.

\subsubsection{IIB flux compactifications}
\label{sec:D4IIB}

We now want to consider gaugings with an origin in type IIB supergravity.
$N=4$ supergravity can be obtained by an orientifold compactification of IIB \cite{Frey:2002hf,Kachru:2002he}
and in the simplest $T^6/\mathbbm{Z}_2$ case this yields the ungauged theory with $n=6$, i.e. the global symmetry group is
$G = {\rm SL}(2) \times {\rm SO}(6,6)$. Here, the ${\rm SL}(2)$ factor is the symmetry that was already present
in ten dimensions and ${\rm SO}(6,6)$ contains the ${\rm GL}(6)$ symmetry group associated with the torus $T^6$.
The compactification thus
yields the theory in a symplectic frame in which ${\rm SL}(2) \times {\rm GL}(6)$ is realized off-shell. Turning on fluxes
results in gaugings of the theory that are purely electric in this particular symplectic frame.
This is the class of gaugings to be examined in this subsection.

An ${\rm SO}(6,6)$ vector decomposes under ${\rm GL}(6)={\rm U}(1) \times {\rm SL}(6)$ into
${\bf 6} \oplus {\bf \overline 6}$.
The vector fields $A_\mua{}^{\Ma\aa}$ split accordingly into electric
ones $A_\mua{}^{\La\aa}$ and magnetic ones $A_{\mua\,\La}{}^{\aa}$ where $\La=1 \ldots 6$ is a (dual) ${\rm SL}(6)$ vector index.
The ${\rm SO}(6,6)$ metric takes the form
\begin{align}
  \eta_{\Ma\Mb} &= \begin{pmatrix} \eta_{\La\Lb} & \eta_\La{}^\Lb \\ \eta^\La{}_\Lb & \eta^{\La\Lb} \end{pmatrix}
                 = \begin{pmatrix} 0 & \delta_\La^\Lb \\ \delta^\La_\Lb & 0 \end{pmatrix} \; .
\end{align}   
The gauge group generators \eqref{DefXD4} split as $X_{\Ma\aa}=(X_{\La\aa}, \, X^\La{}_\aa)$
and a purely electric gauging satisfies $X^\La{}_\aa=0$.
The tensors $\xi_{\aa\Ma}$ and $f_{\aa\Ma\Mb\Mc}$ decompose into the following
representations
\begin{align}
  ({\bf 2},{\bf 12}) \, & \rightarrow \,  ({\bf 2},{\bf 6}) \oplus ({\bf 2},{\bf \overline 6}) \, , \nonumber \\
  ({\bf 2},{\bf 220}) \, & \rightarrow \, 
     ({\bf 2},{\bf 6}) \oplus ({\bf 2},{\bf 20}) \oplus ({\bf 2},{\bf 84}) \oplus ({\bf 2},{\bf \overline{84}})
     \oplus ({\bf 2}, {\bf \overline{20}}) \oplus ({\bf 2},{\bf \overline 6}) \; . 
\end{align}     
From \eqref{DefXD4} one finds that the condition $X^\La{}_\aa=0$ demands most of these components to vanish,
only the $({\bf 2},{\bf 20})$ and
a particular combinations of the two $({\bf 2},{\bf 6})$'s are allowed to be non-zero.
Explicitly we find for the general electric gaugings in this frame
\begin{align}
   \xi_{\aa\Ma} &= (\xi_{\aa\La},\, \xi_\aa{}^\La) = (\xi_{\aa\La},\,0) \, , \nonumber \\
   f_{\aa\Ma\Mb\Mc} &= ( f_{\aa\La\Lb\Lc}, \, f_{\aa\La\Lb}{}^\Lc, \, f_{\aa\La}{}^{\Lb\Lc}, \, f_\aa{}^{\La\Lb\Lc} )
                    = ( f_{\aa\La\Lb\Lc}, \, \xi^{\phantom{\Lc}}_{\aa[\La} \delta_{\Lb]}^\Lc, \, 0, \, 0 ) \; .
   \label{IIBgaugings}		    
\end{align}
This Ansatz automatically satisfies most of the quadratic constraints \eqref{QConD4}, the only consistency constraint
left is
\begin{align}
   f_{(\aa[\La\Lb\Lc} \, \xi_{\ab)\Ld]} &= 0 \; .
   \label{fxi}
\end{align}
Thus for $\xi_{\aa\La} = 0$ we find $f_{\aa\La\Lb\Lc}$ to be unconstrained, i.e. every choice of $f_{\aa\La\Lb\Lc}$
gives a valid gauged theory. It turns out that $f_{\aa\La\Lb\Lc}$ corresponds to the possible three-form fluxes that can
be switched on. These theories and extensions of them
were already described and analyzed in \cite{D'Auria:2002tc,D'Auria:2003jk}.
It was noted in \cite{deWit:2003hq} that not all $N=4$ models that come from $T^6/\mathbbm{Z}_2$ orientifold
compactifications can be embedded
into the $N=8$ models from torus reduction of IIB, since for the latter the fluxes have to satisfy the constraint
$f_{\aa[\La\Lb\Lc} f_{\ab\Ld\Le\Lf]} = 0$.

Searching for solutions to the constraint \eqref{fxi} with $\xi_{\aa\La}$ non-vanishing one finds that
the possible solutions have the form
\begin{align}
   f_{\aa\La\Lb\Lc} &= \xi_{\aa[\La} \, A_{\Lb\Lc]} \; , && \text{or} &
   f_{\aa\La\Lb\Lc} &= \epsilon^{\ab\ac} \,  B_{\aa[\La} \, \xi_{\ab\Lb} \,  \xi_{\ac\Lc]} \; ,
\end{align}
with unconstraint $\xi_{\aa\La}$, $A_{\La\Lb}=A_{[\La\Lb]}$ and $B_{\aa\La}$, respectively.

Theories
with both $f_{\aa\Ma\Mb\Mc}$ and $\xi_{\aa\Ma}$ non-zero were not yet considered in the literature.
For $f_{\aa\Ma\Mb\Mc}=0$ the remaining quadratic constraints on $\xi_{\aa\Ma}$ demands it to be of the form
$\xi_{\aa\Ma} = v_{\aa} \, w_\Ma$, with $v_\aa$ arbitrary and $w_\Ma$ lightlike, i.e. $w_\Ma w^\Ma = 0$.
Thus for vanishing $f_{\aa\Ma\Mb\Mc}$ the solution for $\xi_{\aa\Ma}$ is unique up to ${\rm SL}(2) \times {\rm SO}(6,n)$
transformations. This solution corresponds to the gauging that can be obtained from Scherk-Schwarz reduction from $D=5$ with
a non-compact ${\rm SO}(1,1)$ twist, which was constructed in \cite{Villadoro:2004ci} for the case of one vector multiplet.
This suggests that in certain cases non-vanishing $\xi_{\aa\Ma}$ corresponds to torsion on the internal manifold. But this does
not apply for the IIB reductions in this section since $\xi_{\aa\La}$ is a doublet under the global ${\rm SL}(2)$ symmetry of IIB,
while a torsion parameter should be a singlet.
We have shown that these theories with non-vanishing $\xi_{\aa\La}$ are consistent $N=4$ supergravities,
but their higher-dimensional origin remains to be elucidated.

The list of gauged $N=4$ supergravities that were presented in this section is, of course, far from complete.
One could, for example, discuss other orientifold compactifications of IIA and IIB supergravity, for all of which turning
on fluxes yields gauged theories in four dimensions \cite{Angelantonj:2003rq,Angelantonj:2003up}.
However, the examples discussed were hopefully representative enough to show that indeed all the various gaugings
appearing in the literature can be embedded in the universal formulation presented above. 
New classes of gaugings are those with both $f_{\aa\Ma\Mb\Mc}$ and $\xi_{\aa\Ma}$ non-vanishing.
Every solution of the quadratic constraints \eqref{QConD4}
yields a consistent gauging . For additional examples see \cite{Jonas:Thesis}.

\section{Gauged $N=4$ supergravities in $D=5$}
\setcounter{equation}{0}
\label{sec:D5}

In analogy to the four dimensional theory presented in the last section
we now describe the general gauged $N=4$ (half-maximal) supergravity in five spacetime dimensions\footnote{
Sometimes the half-maximal supergravities in $D=5$ are referred to as $N=2$ theories. We prefer the notation $N=4$
since they are related to the $N=4$ theories in four dimensions via a torus reduction.
In this notation the minimal supergravity in $D=5$ is denoted as $N=2$.}.
The general gauging in $D=5$ is parameterized by three real tensors $f_{\Ma\Mb\Mc}$, $\xi_{\Ma\Mb}$ and $\xi_{\Ma}$, taking
the role of $f_{\aa\Ma\Mb\Mc}$ and $\xi_{\aa\Ma}$ from the last section.
Our presentation is less detailed than for the four dimensional theory because for the case $\xi_\Ma=0$ 
these theories were already presented in the literature \cite{Dall'Agata:2001vb}.
On the other hand, gaugings with vanishing $f_{\Ma\Mb\Mc}$ and $\xi_{\Ma\Mb}$ but non-zero $\xi_\Ma$
have a non-semi-simple gauge group and
originate in generalized dimensional reduction from $D=6$ supergravity \cite{Villadoro:2004ci}.
Here we complete the analysis of \cite{Villadoro:2004ci,Dall'Agata:2001vb} by including 
gaugings with all tensors $f_{\Ma\Mb\Mc}$, $\xi_{\Ma\Mb}$ and $\xi_{\Ma}$ non-zero. 
We give the complete bosonic Lagrangian and Killing spinor equations and at the end of this section
make contact with the four dimensional theory.

\subsection{Quadratic constraints and gauge algebra}
\label{sec:D5qu}

The global symmetry group of ungauged $D=5$, $N=4$ supergravity is $G={\rm SO}(1,1) \times {\rm SO}(5,n)$,
where $n \in \mathbbm{N}$ counts the number of vector multiplets. The theory contains Abelian vector gauge fields
that form one vector $A_\mua^\Ma$ and one scalar $A_\mua^\Nv$ under ${\rm SO}(5,n)$.
Note that the index $\Ma=1 \ldots 5+n$ now is a vector index of ${\rm SO}(5,n)$ while in the last
section we used it for ${\rm SO}(6,n)$. The vector fields carry ${\rm SO}(1,1)$ charges $1/2$ and $-1$, respectively, i.e.
\begin{align}
   \delta_\Na A_\mua^\Ma &= \frac 1 2 A_\mua^\Ma \; , &
   \delta_\Na A_\mua^\Nv &= - A_\mua^\Nv \; ,
\end{align}
where $\delta_\Na$ denotes the ${\rm SO}(1,1)$ action. The corresponding algebra generator is denoted $t_\Na$ while
the ${\rm SO}(5,n)$ generators are $t_{\Ma\Mb}=t_{[\Ma\Mb]}$. For the representations of the vector gauge fields these
generators explicitly read
\begin{align}
   {t_{\Ma\Mb \, \Mc}}^\Md &= \delta^\Md_{[\Ma} \eta^{\phantom{\Md}}_{\Mb]\Mc} \; , &
   {t_{\Na\Ma}}^\Mb &= - \frac 1 2 \delta_\Ma^\Mb \; , &
   {t_{\Ma\Mb \, \Nv}}^\Nv &= 0 \; , &
   {t_{\Na\Nv}}^\Nv &= 1 \; .
   \label{tvf}
\end{align}
The general gauging of the theory is parameterized by tensors $f_{\Ma\Mb\Mc}=f_{[\Ma\Mb\Mc]}$, $\xi_{\Ma\Mb}=\xi_{[\Ma\Mb]}$
and $\xi_\Ma$. They designate the gauge group and assign the vector gauge fields to the gauge group generators. The general
covariant derivative reads
\begin{align}
   D_\mua &= \nabla_\mua - g \, A_\mua^\Ma \, f_\Ma{}^{\Mb\Mc} \, t_{\Mb\Mc}
			 - g \, A_\mua^\Nv \, \xi^{\Mb\Mc} \, t_{\Mb\Mc}
                         - g \, A_\mua^\Ma \, \xi^\Mb \, t_{\Ma\Mb}  
                         - g \, A_\mua^\Ma \, \xi_\Ma \, t_\Na \; ,
  \label{CovDivD5}			 
\end{align}
where the indices are raised and lowered by using the ${\rm SO}(5,n)$ metric $\eta_{\Ma\Mb}$ and $g$ is the gauge
coupling constant.
In order that the above expression is $G$ invariant we need $f_{\Ma\Mb\Mc}$ and $\xi_\Ma$ to carry ${\rm SO}(1,1)$
charge $-1/2$ and $\xi_{\Ma\Mb}$ to have charge $1$.
By $G$ invariance we mean a formal invariance treating the
$f_{\Ma\Mb\Mc}$, $\xi_{\Ma\Mb}$ and $\xi_\Ma$ as spurionic objects that transform under $G$.
However, as soon as we choose particular values for these tensors the global $G$
invariance is broken and only a local $G_0 \subset G$ invariance is left.

To guarantee the closure of the gauge group and the consistency of the gauging we need the following
quadratic constraints to be satisfied for a general gauging
\begin{align}
   \xi_\Ma \xi^\Ma &= 0 \; , &
   \xi_{\Ma\Mb} \xi^\Mb &= 0 \; , &
   f_{\Ma\Mb\Mc} \xi^\Mc &= 0 \; , \nonumber \\
   3 f_{\Me[\Ma\Mb} \, f_{\Mc\Md]}{}^\Me &= 2 f_{[\Ma\Mb\Mc} \, \xi_{\Md]} \; ,  &
   {\xi_{\Ma}}^\Md \, f_{\Md\Mb\Mc} &= \xi_\Ma \, \xi_{\Mb\Mc} - \xi_{[\Mb} \, \xi_{\Mc]\Ma} \; .
   \label{XiRel}
\end{align}
This implies for example that $\xi_\Ma$ has to vanish for $n=0$ since for an Euclidean metric $\eta_{\Ma\Mb}$ one 
has no lightlike vectors. In general, however, all three tensors may be non-zero at the same time.
For the sake of discussing the closure of the gauge group it is convenient to consider the group action on the vector field
representation defined by \eqref{tvf}. Introducing the composite index $\cMa=\{ \Nv, \, \Ma\}$, i.e.
$A_\mua^\cMa = (A_\mua^\Nv, \, A_\mua^\Ma)$, we have
\begin{align}
   D_\mua \, \Lambda^\cMa &= \nabla_\mua \, \Lambda^\cMa + g \, A_\mua^\cMb \, X_{\cMb\cMc}{}^\cMa  \, \Lambda^\cMc \; ,
\end{align}
where the gauge group generators $X_{\cMa\cMb}{}^\cMc = (X_{\cMa})_\cMb{}^\cMc$ are given by
\begin{align}
   {X_{\Ma\Mb}}^\Mc &= - f_{\Ma\Mb}{}^\Mc - \frac 1 2 \eta_{\Ma\Mb} \xi^\Mc + \delta_{[\Ma}^\Mc \xi_{\Mb]} \; , &
   {X_{\Ma\Nv}}^\Nv &= \xi_\Ma \; , &
   {X_{\Nv\Ma}}^\Mb &= - {\xi_\Ma}^\Mb  \; , &
   \label{GenD5}
\end{align}
and all other components vanish. For the commutator of these generators one finds
\begin{align}
   [ X_\cMa , X_\cMb ] &= - {X_{\cMa\cMb}}^\cMc \, X_\cMc \; ,
   \label{ClosureD5}
\end{align}
i.e. the gauge group closes and ${X_{\cMa\cMb}}^\cMc$ itself takes the role of a generalized structure constant.
The closure relation \eqref{ClosureD5} is equivalent to the quadratic constraint \eqref{XiRel}.

For gaugings with only $f_{\Ma\Mb\Mc}$ non-zero we see that this tensor is a structure constant for a subgroup $G_0$
of ${\rm SO}(5,n)$ that is gauged by using $A_\mua^\Ma$ as vector gauge fields.
If only $\xi_{\Ma\Mb}$ is non-zero we find a one-dimensional subgroup of ${\rm SO}(5,n)$
to be gauged with gauge field $A_\mua^\Nv$. And for gaugings with only $\xi_\Ma$ non-zero one finds a $4+n$ dimensional
gauge group ${\rm SO}(1,1) \ltimes {\rm SO}(1,1)^{3+n}$ where the first factor involves the ${\rm SO}(1,1)$ of $G$.

Note further that \eqref{ClosureD5} is of precisely the same form as the closure relation \eqref{ClosureD4} which
we had in four dimensions. This is not by accident but we have just applied a general method of how to gauge
supersymmetric theories which goes under the name of the embedding tensor
\cite{Nicolai:2000sc,deWit:2002vt,deWit:2004nw,Samtleben:2005bp,deWit:2005ub}.
In this language the tensors $f_{\Ma\Mb\Mc}$,
$\xi_{\Ma\Mb}$ and $\xi_\Ma$ are components of the embedding tensor which is a linear map from the vector space of
vector gauge fields to the Lie algebra of invariances of the ungauged theory. Independent of the number of supersymmetries or of the
spacetime dimension the embedding tensor always has to satisfy the quadratic constraint \eqref{ClosureD5}. In addition it
always satisfies a linear constraint which involves extra objects and whose form depends on the number of spacetime dimensions.
For example in $D=4$ the linear constraint involves the antisymmetric tensor $\Omega_{\sMa\sMb}$ and has the form
${X_{(\sMa\sMb}}^\sMd \Omega_{\sMc)\sMd} = 0$ \cite{deWit:2005ub} while in $D=5$ it involves the tensor
$d_{\cMa\cMb\cMc}$ and takes the form \eqref{linConD5} below.
Our presentation of the five dimensional theory is to a large extend based on \cite{deWit:2004nw} where the
corresponding maximal supergravity was presented.

To give the Lagrangian and the gauge transformations of the theory in the next section it is useful to introduce the tensors
$d_{\cMa\cMb\cMd}=d_{(\cMa\cMb\cMd)}$ and $Z^{\cMa\cMb}=Z^{[\cMa\cMb]}$ as follows
\begin{align}
   d_{\Nv\Ma\Mb} &= d_{\Ma\Nv\Mb} = d_{\Ma\Mb\Nv} = \eta_{\Ma\Mb} \; , \qquad \text{all other components zero,}
\end{align}
and
\begin{align}
   Z^{\Ma\Mb} &= \, \ft 1 2 \, \xi^{\Ma\Mb} \; , &
   Z^{\Nv\Ma} &= - Z^{\Ma\Nv} = \, \ft 1 2 \, \xi^\Ma \; .
   \label{ZD5}
\end{align}
The embedding tensor then satisfies
\begin{align}
   {X_{(\cMa\cMb)}}^\cMc &= d_{\cMa\cMb\cMd} Z^{\cMc\cMd} \; .
   \label{linConD5}
\end{align}

\subsection{The general Lagrangian}
\label{sec:D5lag}

We have already introduced the vector fields $A_\mua^\Ma$ and $A_\mua^\Nv$. In addition the bosonic field content
consists of scalars that form the coset ${\rm SO}(1,1) \times {\rm SO}(5,n)/{\rm SO}(5) \times {\rm SO}(n)$ and
two-form gauge fields $B_{\mua\mub\,\cMa} = (B_{\mua\mub \, \Ma}, B_{\mua\mub \, \Nv})$. In the ungauged theory these
two-form fields do not appear in the Lagrangian but can be introduced on-shell as the duals of the vector gauge fields.
In the gauged theory we consider both vector and two-form fields as off-shell degrees of freedom, however,
the latter do not have a kinetic term but couple to the vector fields via a topological term such that they turn dual
to the vectors due to their own equations of motion \cite{deWit:2004nw}. This is analogous to the four dimensional case
where the two-forms turned out to be dual to scalars via the equations of motion.

The ${\rm SO}(1,1)$ part of the scalar manifold is simply described by one real field $\Sigma$ that is a singlet under
${\rm SO}(5,n)$ and carries ${\rm SO}(1,1)$ charge $-1/2$. In addition we have the coset
${\rm SO}(5,n)/{\rm SO}(5) \times {\rm SO}(n)$ which is parameterized by a coset representative
${\cal V}=({\cal V}_\Ma{}^\ya, \, {\cal V}_\Ma{}^\xa)$, where $\ya=1\ldots 5$ and $\xa=1\ldots n$ are
${\rm SO}(5)$ and ${\rm SO}(n)$ vector indices. Our conventions for ${\cal V}$ here are the same as for the 
${\rm SO}(6,n)/{\rm SO}(6) \times {\rm SO}(n)$ coset representative we had in four dimensions, see equations
\eqref{DefEta}, \eqref{CosetSO6n}, \eqref{DefMVV}. In addition to the symmetric matrix $M_{\Ma\Mb}={\cal V} {\cal V}^T$
and its inverse $M^{\Ma\Mb}$ we need the completely antisymmetric
\begin{align}
   M_{\Ma\Mb\Mc\Md\Me} &= \epsilon_{\ya\yb\yc\yd\ye}
                             {{\cal V}_\Ma}^{\ya} {{\cal V}_\Mb}^{\yb} {{\cal V}_\Mc}^{\yc} 
                             {{\cal V}_\Md}^{\yd} {{\cal V}_\Me}^{\ye} \; .
\end{align}

The two-form gauge fields transform dual to the vector gauge field under $G$, i.e. $B_{\mua\mub \, \Ma}$ is
a vector with ${\rm SO}(1,1)$ charge $-1/2$ and $B_{\mua\mub \, \Nv}$ is a singlet carrying charge $1$. 
They enter into the covariant field strength of the vector fields as follows
\begin{align}
   {\cal H}_{\mua\mub}^\cMa &\equiv 2 \partial_{[\mua} A^\cMa_{\mub]} + g {X_{\cMb\cMc}}^\cMa A^\cMb_\mua A^\cMc_\mub
                               + g Z^{\cMa\cMb} B_{\mua\mub\,\cMb} \; .
\end{align}

We now have all objects to give the bosonic Lagrangian of the general gauged $N=4$ supergravity in five dimensions
\begin{align}
   {\cal L}_{\text{bos}} &= {\cal L}_{\text{kin}} + {\cal L}_{\text{top}} + {\cal L}_{\text{pot}} \; .   
\end{align}
It consists of a kinetic part
\begin{align}
   e^{-1} {\cal L}_{\text{kin}} &= \ft 1 2 \, R 
               - \, \ft 1 4 \, \Sigma^2 \, M_{\Ma\Mb} \, {\cal H}^\Ma_{\mua\mub} \, {\cal H}^{\Mb\, \mua\mub}
               - \, \ft 1 4 \, \Sigma^{-4} \, {\cal H}^\Nv_{\mua\mub} \, {\cal H}^{\Nv\,\mua\mub}
	       \nonumber \\ & \qquad
	       - \, \ft 3 2 \, \Sigma^{-2} \, (D_\mua \Sigma)^2 
	       + \, \ft 1 {16} \, (D_\mua M_{\Ma\Mb}) (D^\mua M^{\Ma\Mb}) \; ,
\end{align}
a topological part \cite{deWit:2004nw}
\begin{align}
   {\cal L}_{\text{top}} &= - \frac e {8 \sqrt{2}} \epsilon^{\mua\mub\muc\mud\mue} 
    \bigg\{ g Z^{\cMa\cMb} B_{\mua\mub \, \cMa} \left[ D_\muc B_{\mud\mue \, \cMb}
                   + 4 d_{\cMb\cMc\cMd} A^\cMc_{[\muc} \Big( \partial_\mud A_{\mue]}^\cMd 
	                      + \ft 1 3 g {X_{\cMe\cMf}}^\cMc A_\mud^\cMe A_{\mue]}^\cMf \Big) \right]
			      \nonumber \\ & \qquad \qquad \qquad \quad
      - \ft 8 3 \, d_{\cMa\cMb\cMc} \, A_\mua^\cMa \, \partial_\mub A_\muc^\cMb \, \partial_\mud A_\mue^\cMc
      - 2 \, g \, d_{\cMa\cMb\cMc} \, {X_{\cMd\cMe}}^\cMa \, A_\mua^\cMb \, A_\mub^\cMd \, A_\muc^\cMe \, \partial_\mud A_\mue^\cMc
		      \nonumber \\ & \qquad \qquad \qquad \qquad
      - \ft 2 5 \, g^2 \, d_{\cMa\cMb\cMc} \, {X_{\cMd\cMe}}^\cMa \, {X_{\cMf\cMg}}^\cMc  \,
              A_\mua^\cMb \, A_\mub^\cMd \, A_\muc^\cMe \, A_\mud^\cMf \, A_\mue^\cMg \bigg\} \; ,
\end{align}
and a scalar potential
\begin{align}
   e^{-1} {\cal L}_{\text{pot}} &= -g^2 V \nonumber \\
    &= - \frac {g^2} {4} \Big[
       \xi_{\Ma\Mb\Mc} \xi_{\Md\Me\Mf} \Sigma^{-2} \left(
       \ft 1 {12} M^{\Ma\Md} M^{\Mb\Me} M^{\Mc\Mf} 
      -\ft 1 {4} M^{\Ma\Md} \eta^{\Mb\Me} \eta^{\Mc\Mf} 
      +\ft 1 {6} \eta^{\Ma\Md} \eta^{\Mb\Me} \eta^{\Mc\Mf} \right)
      \nonumber \\ & \qquad \qquad
      +\ft 1 4 \xi_{\Ma\Mb} \xi_{\Mc\Md} \Sigma^4 \left( M^{\Ma\Mc} M^{\Mb\Md} - \eta^{\Ma\Mc} \eta^{\Mb\Md} \right)
      +\xi_\Ma \xi_\Mb \Sigma^{-2} M^{\Ma\Mb}
      \nonumber \\ & \qquad \qquad \qquad
      +\ft 1 3 \sqrt{2} \xi_{\Ma\Mb\Mc} \xi_{\Md\Me} \Sigma M^{\Ma\Mb\Mc\Md\Me} \Big] \; .
   \label{PotD5}      
\end{align}
For $\xi_{\Ma}=0$ the latter agrees with the potential given in \cite{Dall'Agata:2001vb}.

The topological term seems complicated, but its variation with respect to the vector and tensor gauge fields
takes a rather simple and covariant form, namely
\begin{align}
   \delta {\cal L}_{\text{top}} &= \frac{e} {4 \sqrt{2}} \epsilon^{\mua\mub\muc\mud\mue}
   \left( \ft 1 3 \, g \, Z^{\cMa\cMb} \,  {\cal H}^{(3)}_{\mua\mub\muc \, \cMa} \, \Delta B_{\mud\mue \, \cMb}
          + d_{\cMa\cMb\cMc} \, {\cal H}_{\mua\mub}^\cMa \, {\cal H}_{\muc\mud}^\cMb \, \delta A_\mue^\cMc \right)
   + \text{tot. deriv.}	   \; ,
   \label{VaryLtop}   
\end{align}
where we have used the covariant variation
\begin{align}
  \Delta B_{\mua\mub \, \cMb} 
      &\equiv Z^{\cMa\cMb} \left( \delta B_{\mua\mub \, \cMb} 
         - 2 d_{\cMb\cMc\cMd} A_{[\mua}^\cMc \delta A_{\mub]}^\cMd \right) \; ,
\end{align}
and the covariant field strength of the two-form gauge fields
\begin{align}
   Z^{\cMa\cMb} {\cal H}^{(3)}_{\mua\mub\muc \, \cMb} 
     &=  Z^{\cMa\cMb} \left[ 3 \, D_{[\mua} \, B_{\mub\muc] \cMb} 
             + 6 \, d_{\cMb\cMc\cMd} \, A^\cMc_{[\mua}  \left( \partial^{\phantom{\cMd}}_\mub A_{\muc]}^\cMd 
	                      + \ft 1 3 \, g \, {X_{\cMe\cMf}}^\cMd \, A_\mub^\cMe \, A_{\muc]}^\cMf \right) \right] \; .
\end{align}
Note that the two-forms appear in the Lagrangian always projected with $Z^{\cMa\cMb}$, i.e. they completely decouple from the
theory for the ungauged case $g \rightarrow 0$, but also for the gauged theory there are never all two-forms entering the
Lagrangian. For gaugings with only $f_{\Ma\Mb\Mc}$ non-zero we have $Z^{\cMa\cMb}=0$ and thus no two-forms are needed.
In the last equation we also defined the field strength of the two-forms only under $Z^{\cMa\cMb}$ projection because
only then it transforms covariantly under the following gauge transformations \cite{deWit:2004nw}
\begin{align}
   \delta A_\mua^\cMa &= D_\mua \Lambda^\cMa - g Z^{\cMa\cMb} \Xi_{\mua \, \cMb} \; , \nonumber \\
   \Delta B_{\mua\mub \, \cMa} 
      &= \left( 2 D_{[\mua} \Xi_{\mub]\cMa} - 2 d_{\cMa\cMb\cMc} {\cal H}_{\mua\mub}^\cMb \Lambda^\cMc \right)  \;.
    \label{D5gaugetrafo}
\end{align}
Here $\Lambda^\cMa=\Lambda^\cMa(x)$ and $\Xi_{\mua \, \cMa}=\Xi_{\mua \, \cMa}(x)$ parameterize the (tensor) 
gauge transformations. Also the field strength ${\cal H}^{\cMa}_{\mua\mub}$ transforms covariantly under these
transformations, i.e. 
\begin{align}
   \delta {\cal H}_{\mua\mub}^\cMa &= - g \Lambda^\cMb {X_{\cMb\cMc}}^\cMa {\cal H}_{\mua\mub}^\cMc \;.
\end{align}
The topological term ${\cal L}_{\text{top}}$ is invariant under \eqref{D5gaugetrafo} up to a total derivative.
The algebra of gauge transformations closes analogous to the one we found in four dimensions \eqref{D4gaugealgebra}.

Varying the two-forms in the Lagrangian yields the equation of motion
\begin{align}
   Z^{\cMa\cMb} \left( \frac{1} {6 \sqrt{2}} \epsilon_{\mua\mub\muc\mud\mue} \,  {\cal H}^{(3)\,\muc\mud\mue}_{\cMb} 
                      - {\cal M}_{\cMb\cMc} {\cal H}_{\mua\mub}^{\cMc} \right) &= 0 \; ,
  \label{Dual23}		      
\end{align}
where we have used
\begin{align}
   {\cal M}_{\cMa\cMb} &\equiv \begin{pmatrix} \Sigma^{-4} & 0 \\
                                          0 & \Sigma^2 M_{\Ma\Mb} \end{pmatrix} \; .
\end{align}
Due to equation \eqref{Dual23} the two-forms become dual to the vector gauge fields as was announced earlier.

\subsection{Killing spinor equations}
\label{sec:D5kill}

We now turn to the fermions of the five dimensional theory in order to give the Killing spinor equations.
The fermions come in representations of the maximal compact subgroup $H={\rm USp}(4) \times {\rm SO}(n)$ of $G$,
where ${\rm USp}(4)$ is the covering group of ${\rm SO}(5)$.
In the gravity multiplet there are four gravitini $\psi_{\mua\ja}$ and four spin $1/2$ fermions $\chi_\ja$, both
vectors under ${\rm USp}(4)$ and singlets under ${\rm SO}(n)$, $i=1\ldots 4$. In the $n$ vector multiplets there are $4n$
spin $1/2$ fermions $\lambda^\xa_\ja$ which form a vector under both ${\rm USp}(4)$ and ${\rm SO}(n)$, $\xa=1\ldots n$.
All fermions are pseudo-Majorana, i.e. they satisfy a pseudo-reality constraint of the form
$\xi_\ja = \Omega_{\ja\jb} C ({\bar \xi}^\jb)^T$, where $\Omega_{\ja\jb}$ is the ${\rm USp}(4)$ invariant symplectic form
and $C$ is the charge conjugation matrix.

The coset representative ${\cal V}_\Ma{}^\ya$ transforms as a ${\bf 5}$ under ${\rm USp}(4)$ and can alternatively
be expressed as ${\cal V}_\Ma{}^{\ja\jb}={\cal V}_\Ma{}^{[\ja\jb]}$ subject to
\begin{align}
   {\cal V}_\Ma{}^{\ja\jb} \Omega_{\ja\jb} &= 0 \; , &
   ( {\cal V}_\Ma{}^{\ja\jb} )^* &= \Omega_{\ja\jc} \Omega_{\jb\jd} {\cal V}_\Ma{}^{\jc\jd} \; .
\end{align}
Under supersymmetry transformations parameterized by $\epsilon_\ja=\epsilon_\ja(x)$ we have
\begin{align}
   \delta \psi_{\mua\ja}  &= D_\mua \epsilon_\ja 
      - \frac i 6 \left( \Omega_{\ja\jb} \Sigma {\cal V}_{\Ma}{}^{\jb\jc} {\cal H}_{\mub\muc}^\Ma
                          - \ft 1 4 \sqrt{2} \, \delta_\ja^\jc \Sigma^{-2} {\cal H}_{\mub\muc}^\Nv \right)
		  \left( \Gamma_\mua{}^{\mub\muc} - 4 \delta_\mua^\mub \Gamma^\muc \right) \epsilon_\jc
       \nonumber \\ & \qquad		  
      + \frac {i g} {\sqrt{6}} \, \Omega_{\ja\jb} \, A_1^{\jb\jc} \, \Gamma_\mua  \, \epsilon_\jc \; ,
   \nonumber \\
   \delta \chi_\ja &= - \ft 1 2 \, \sqrt{3} \, i \, (\Sigma^{-1} D_\mua \Sigma) \, \Gamma^\mua \epsilon_\ja
                      - \ft 1 6 \, \sqrt{3} \, 
		      \left( \Sigma \, \Omega_{\ja\jb} \, {\cal V}_{\Ma}{}^{\jb\jc} {\cal H}_{\mua\mub}^\Ma
            + \ft 1 2 \sqrt{2} \, \Sigma^{-2} \, \delta_\ja^\jc \, {\cal H}_{\mua\mub}^\Nv \right) \Gamma^{\mua\mub} \epsilon_\jc
        \nonumber \\ & \qquad			     
		     + \sqrt{2} \, g \, \Omega_{\ja\jb} \, A_2^{\jc\jb} \, \epsilon_\jc \; ,
   \nonumber \\
  \delta \lambda_\ja^\xa &= i \, \Omega^{\jb\jc} \, ( {\cal V}_{\Ma}{}^{\xa} D_\mua {\cal V}_{\ja\jb}{}^\Ma ) \Gamma^\mua \epsilon_\jc
                       - \ft 1 4 \, \Sigma \, {\cal V}_{\Ma}{}^\xa \, {\cal H}_{\mua\mub}^\Ma \, \Gamma^{\mua\mub} \, \epsilon_\ja
		       + \sqrt{2} \, g \, \Omega_{\ja\jb} \, A_2^{\xa\jc\jb} \, \epsilon_\jc	      	      \; .
\end{align}
Here we have neglected higher order fermion terms. These fermion variations could formally be
read off from \cite{Dall'Agata:2001vb}. But the fermion shift matrices
$A_{1 \ja\jb}$, $A_{2 \ja\jb}$ and $A_{2 \ja\jb}^\xa$ which are defined below now include contributions
from the vector $\xi_\Ma$.

Using ${\cal V}_\Ma{}^\xa$ and ${\cal V}_\Ma{}^{\ja\jb}$ we can define 
from $f_{\Ma\Mb\Mc}$, $\xi_{\Ma\Mb}$ and $\xi_\Ma$
scalar dependent tensors that transform under $H$. The vector $\xi_\Ma$ gives
\begin{align}
   \tau^{\ja\jb} &= \Sigma^{-1} {{\cal V}_\Ma}^{\ja\jb} \, \xi^\Ma  \; ,&
   \tau^\xa &= \Sigma^{-1} {{\cal V}_\Ma}^\xa \, \xi^\Ma \; ,
\end{align}
from the 2-form $\xi_{\Ma\Mb}$ one gets
\begin{align}   
   \zeta^{\ja\jb} &= \sqrt{2} \, \Sigma^2 \Omega_{\jc\jd} \, {{\cal V}_\Ma}^{\ja\jc} {{\cal V}_\Mb}^{\jb\jd} \, \xi^{\Ma\Mb}  \; ,&
   \zeta^{\xa\ja\jb} &= \Sigma^2 {{\cal V}_\Ma}^{\xa} {{\cal V}_\Mb}^{\ja\jb} \, \xi^{\Ma\Mb} \; ,
\end{align}
and the 3-form $f_{\Ma\Mb\Mc}$ yields
\begin{align}   
   \rho^{\ja\jb} &= - \ft 2 3 \, \Sigma^{-1} {{\cal V}^{\ja\jc}}_{\Ma} {{\cal V}^{\jb\jd}}_{\Mb} {{\cal V}^\Mc}_{\jc\jd} \, {f^{\Ma\Mb}}_\Mc \; ,
   &
   \rho^{\xa\ja\jb} &= \sqrt{2} \, \Sigma^{-1} \, \Omega_{\jc\jd} \,
                          {{\cal V}_\Ma}^{\xa} {{\cal V}_\Mb}^{\ja\jc} {{\cal V}_\Mc}^{\jb\jd} \, f^{\Ma\Mb\Mc}  \; ,
\end{align}
where $\lambda^{\ja\jb} = \lambda^{[\ja\jb]}$, $\zeta^{\ja\jb} = \zeta^{(\ja\jb)}$, $\zeta^{\xa\ja\jb} = \zeta^{\xa[\ja\jb]}$,
$\rho^{\ja\jb} = \rho^{(\ja\jb)}$, $\rho^{\xa\ja\jb} = \rho^{\xa(\ja\jb)}$.
\footnote{
Our notation translates into that of \cite{Dall'Agata:2001vb} as follows:
$a_\mua = A_\mua^\Nv$, $\Lambda^\Ma_\Mb = \frac {g} {g_A}  {\xi^\Ma}_\Mb$, $f_{\Ma\Mb}^\Mc = - \frac g {g_S}  f_{\Ma\Mb}{}^\Mc$,
$U_{\ja\jb} = - \frac {g} {6 g_A} \zeta_{\ja\jb}$, $V_{\ja\jb}^\xa = - \frac {g} {\sqrt{2} g_A} \zeta_{\ja\jb}^{\xa}$,
$S_{\ja\jb} = \frac g {3 g_S}  \rho_{\ja\jb}$,
$T^\xa_{\ja\jb} = \frac g {\sqrt{2} g_S}  \rho^\xa_{\ja\jb}$.
}
The fermion shift matrices can now be defined as
\begin{align}
   A_1^{\ja\jb} &= \frac 1 {\sqrt{6}} \left( - \zeta^{\ja\jb} 
                                        + 2 \rho^{\ja\jb} \right) \, , \nonumber \\
   A_2^{\ja\jb} &= \frac 1 {\sqrt 6} \left( \zeta^{\ja\jb} 
                                     + \rho^{\ja\jb} 
				     + \ft 3 {2} \, \tau^{\ja\jb} \right) \, , \nonumber \\
   A_2^{\xa\ja\jb} &= \frac 1 2 \left( - \zeta^{\xa\ja\jb}
                                      + \rho^{\xa\ja\jb}
				      - \, \ft 1 4 \, \sqrt{2} \, \tau^{\xa} \, \Omega^{\ja\jb} \right) \; .
\end{align}
These matrices do not only appear in the fermion variations but also in the fermion mass terms that have to appear
in the Lagrangian of the gauged theory
\begin{align}
   e^{-1} {\cal L}_{\text{f.mass}} \, &= \, \frac {\sqrt{6} \, i \, g} 4 \, \Omega_{\jc\ja} \, A_1^{\ja\jb}  \, 
                 \bar \psi^\jc_\mua \Gamma^{\mua\mub} \psi_{\mub \, \jb}
                   + \sqrt{2} \, g \, \Omega_{\jc\jb} \, A_2^{\jb\ja} \, \bar \psi^\jc_\mua \Gamma^{\mua} \chi_{\ja}
                   + \sqrt{2} \, g \, \Omega_{\jc\jb} \, A_2^{\jb\ja\xa} \, \bar \psi^\jc_\mua \Gamma^{\mua} \lambda_{\ja}^{\xa}		   
    \; .		   
\end{align}
Note that we have only given those terms that involve the gravitini.
Supersymmetry imposes the following condition on the fermion shift matrices
\begin{align}
   \Omega_{\jc\jd} \left( A_1^{\ja\jc} A_1^{\jb\jd} - A_2^{\ja\jc} A_2^{\jb\jd} - A_2^{\xa\ja\jc} A_2^{\xa\jb\jd} \right)
                 &= - \frac 1 4 \Omega^{\ja\jb} V  \; ,
   \label{WardD5}		 
\end{align}
where the scalar potential appears on the right hand side. Again this condition is satisfied as a consequence of the
quadratic constraint \eqref{XiRel}.

\subsection{Dimensional reduction from $D=5$ to $D=4$}
\label{sec:reduction}

Starting from a five dimensional $N=4$ supergravity one can perform a circle reduction
to get a four dimensional $N=4$ supergravity. Thus any five dimensional gauging described by $f_{\Ma\Mb\Mc}$,
$\xi_{\Ma\Mb}$ and $\xi_\Ma$ must give rise to a particular four dimensional gauging characterized by
$f_{\aa\Ma\Mb\Mc}$ and $\xi_{\aa\Ma}$. In other words the set of five dimensional gaugings is embedded into the
set of four dimensional gaugings and we now want to make this embedding explicit.
This yields additional examples of four dimensional gaugings, but it is also interesting
in the context of string dualities in presence of fluxes since the two tensors $f_{\Ma\Mb\Mc}$ and $\xi_{\Ma\Mb}$ in $D=5$
turn out to be parts of the single tensor $f_{\aa\Ma\Mb\Mc}$ under the larger duality group
in $D=4$. Thus, as usual, one gets a more
unified description of gaugings with different higher dimensional origin when compactifying the supergravity
theory further. With all the group structure at hand it is not necessary to
explicitly perform the dimensional reduction but we can read off the connection from the formulas for the
covariant derivatives \eqref{CovDivD4} and \eqref{CovDivD5} (that is from the embedding tensor).

A five dimensional theory with $n$ vector multiplets yields a four dimensional theory with $n+1$ vector multiplets.
One way to understand that is by counting scalar fields. There are $5n+1$ scalars already present in five dimensions and
in addition one gets one scalar from the metric and $6+n$ scalars form the vector fields which gives $6n+8$ in total and agrees
with the number of scalars in the coset ${\rm SL}(2) \times {\rm SO}(6,n+1)/{\rm SO}(2) \times {\rm SO}(6) \times {\rm SO}(n+1)$.
When breaking the ${\rm SO}(6,n+1)$ into ${\rm SO}(1,1)_A \times {\rm SO}(5,n)$
the vector representation splits into an ${\rm SO}(5,n)$ vector $v^\Ma$ and two scalars $v^\oplus$ and $v^\ominus$
with charges $0$, $1/2$ and $-1/2$, respectively,
under ${\rm SO}(1,1)_A$. When breaking the ${\rm SL}(2)$ into ${\rm SO}(1,1)_B$ the vector
splits into two scalars $v^+$ and $v^-$ with charges $1/2$ and $-1/2$ under ${\rm SO}(1,1)_B$. The four dimensional
vector fields therefore split into $A_\mua^{\Ma+}$, $A_\mua^{\Ma-}$, $A_\mua^{\oplus+}$, $A_\mua^{\oplus-}$,
$A_\mua^{\ominus+}$ and $A_\mua^{\ominus-}$. We can now identifying the five dimensional vector fields as
\begin{align}
   A_\mua^{\Ma} &= A_\mua^{\Ma+} \, , &
   A_\mua^\Nv &= A_\mua^{\ominus-} \, ,
   \label{IdentA}
\end{align}
and these fields carry charges $1/2$ and $-1$ under the diagonal of ${\rm SO}(1,1)_A$ and ${\rm SO}(1,1)_B$ and
the five dimensional ${\rm SO}(1,1)$ therefore has to be this diagonal. Thus the five dimensional
global symmetry generators are given in terms of the four dimensional ones as follows
\begin{align}
   t_{\Na} &= t^{{\rm SL}(2)}_{+-} + t^{{\rm SO}(6,n+1)}_{\ominus\oplus} \; , &
   t_{\Ma\Mb} &= t^{{\rm SO}(6,n+1)}_{[\Ma\Mb]} \; .
   \label{IdentT}
\end{align}
The vector fields $A_\mua^{\Ma-}$, $A_\mua^{\oplus+}$ are the four dimensional duals of
$A_\mua^{\Ma+}$ and $A_\mua^{\ominus-}$, they come from the two-form gauge fields in five dimensions.
The vector fields $A_\mua^{\oplus-}$ and $A_\mua^{\ominus+}$ are uncharged under the five dimensional
${\rm SO}(1,1)$, they are the Kaluza-Klein vector coming from the metric and its dual field.

Now, if a four dimensional vector field that was already a vector field in five dimensions \eqref{IdentA}
gauges a four dimensional symmetry that was already a symmetry in five dimensions \eqref{IdentT} the corresponding
gauge coupling in the covariant derivative in $D=4$ has to be the same as in $D=5$.
For the four dimensional covariant derivative \eqref{CovDivD4} one finds
\begin{align}
   D_\mua &= \nabla_\mua - g \, A_\mua{}^{\Ma+} 
                          \left(  \Theta_{+\Ma}{}^{\Mb\Mc} t_{\Mb\Mc}
                           + 2 f_{+\Ma}{}^{\ominus\oplus} t_{\ominus\oplus} 
                           + \xi_{+\Ma} t_{+-} \right)
			    \nonumber \\ & \qquad \quad
                           - g \, A_\mua{}^{\ominus-} 
			  \left(  f_{-\ominus}{}^{\Mb\Mc} t_{\Mb\Mc}
                           + \xi_{-\ominus} t_{\ominus\oplus}
                           - \xi_{-\ominus} t_{+-} \right) \, + D^{\text{add}}_{\mua} \, ,
\end{align}
where $\Theta_{\aa\Ma\Mb\Mc}$ is defined in \eqref{DefThetaF}\footnote{
Note that what we called $n$ in section \ref{sec:D4} 
is now $n+1$ and the index $\Ma$ now is an ${\rm SO}(5,n)$ vector index rather than
a ${\rm SO}(6,n+1)$ index.} and $D^{\text{add}}_{\mua}$ denotes exclusively four dimensional contributions to the covariant derivative.
By comparing with the known covariant derivative in five dimensions \eqref{CovDivD5} one gets
\begin{align}
   \xi_{+\Ma} &= \xi_\Ma \; , &
   f_{+\Ma\oplus\ominus} &= \ft 1 2 \, \xi_\Ma \; , &
   f_{-\ominus\Ma\Mb} &= \xi_{\Ma\Mb} \; , &
   f_{+\Ma\Mb\Mc} &= f_{\Ma\Mb\Mc}  \; . &
\end{align}
For a simple circle reduction it is natural to demand furthermore
$f_{\pm\Ma\Mb\oplus} = 0$, $f_{+\Ma\Mb\ominus}=0$, $f_{-\Ma\Mb\Mc} = 0$,
$f_{-\Ma\oplus\ominus} = 0$, $\xi_{-\Ma}=0$, $\xi_{\pm\oplus}=0$ and $\xi_{\pm\ominus}=0$.
Some of the last quantities, however, may be non-zero for more complicated dimensional reductions
and may then for example correspond to Scherk-Schwarz generators \cite{Villadoro:2004ci}.
But for the ordinary circle reduction we have just given the embedding of the five dimensional gaugings
into the four dimensional ones. In addition to the above equations we have to make sure that
$f_{\aa{\tilde \Ma}{\tilde \Mb}{\tilde \Mc}}$ is totally antisymmetric 
in the last three indices ($\tilde \Ma=\{\Ma,\oplus,\ominus\}$). One can then show that for these
tensors $f_{\aa{\tilde \Ma}{\tilde \Mb}{\tilde \Mc}}$ and $\xi_{\aa{\tilde \Ma}}$ the four dimensional
quadratic constraint \eqref{QConD4}
becomes precisely the five dimensional one \eqref{XiRel} for $f_{\Ma\Mb\Mc}$, $\xi_{\Ma\Mb}$
and $\xi_\Ma$. Also the four and the five dimensional scalar potentials \eqref{VD4}, \eqref{PotD5}
become the same if all scalars that are not yet present in $D=5$ are set to the origin\footnote{
The equality of the scalar potentials is most easily checked at the origin $M=\mathbbm{1}$. If the potentials
do agree there for all possible gaugings the statement is already proven due to the ${\rm SO}(1,1) \times {\rm SO}(5,n)$
covariance of the construction.}.

Due to the antisymmetry of $f_{\aa{\tilde \Ma}{\tilde \Mb}{\tilde \Mc}}$ one finds the following additional terms 
in the $D=4$ covariant derivative:
\begin{align}
   D^{\text{add}}_{\mua}
     &= - g \, A_\mua{}^{\Ma-} \left( 2 {\xi_{\Ma}}^{\Mb} \, t_{\Mb\ominus} + \xi_\Ma \, t_{--} \right)
                \nonumber \\ & \qquad \qquad
                   + g \, A_\mua{}^{\ominus+} \, \xi^{\Mb} \, \left( t_{\Mb\ominus} -  t_{\Mb\oplus} \right)
		   + g \, A_\mua{}^{\oplus+} \, \xi^{\Mb} \left( t_{\Mb\ominus} + t_{\Mb\oplus}   \right) \; .
\end{align}
These are couplings of vector fields to symmetry generators that both only occur in four dimensions.
If one explicitly performs the dimensional reduction by hand these gauge couplings originate from
the dualization of the various fields.

\section{Conclusions}
\setcounter{equation}{0}
\label{sec:conclusions}

The general gaugings of $N=4$ supergravity in $D=5$ and $D=4$
were presented.
The $D=4$ gaugings are parameterized by two ${\rm SL}(2) \times {\rm SO}(6,n)$ tensors
$f_{\aa\Ma\Mb\Mc}$ and $\xi_{\aa\Ma}$, subject to a set of consistency constraints.
New classes of gaugings were found and it was shown how the known gaugings 
are incorporated in this framework. Remarkably, all known examples can be described by turning on only $f_{\aa\Ma\Mb\Mc}$ or
$\xi_{\aa\Ma}$, but we have shown that for a general gauging both tensors can be non-vanishing.
Similarly, in five dimensions the general gaugings
are parameterized by three ${\rm SO}(1,1) \times {\rm SO}(5,n)$ tensors
$f_{\Ma\Mb\Mc}$, $\xi_{\Ma\Mb}$ and $\xi_\Ma$. The gaugings with $\xi_\Ma=0$
were already described in \cite{Dall'Agata:2001vb}, but it is necessary to incorporate $\xi_\Ma$
to also include non-semi-simple gaugings that result from Scherk-Schwarz dimensional reduction \cite{Villadoro:2004ci}.
For a generic gauging all three tensors may be non-zero.
It would be very interesting to understand
how all these gaugings can be obtained
from compactifications of string- or M-theory. 
For example for the $D=4$ gaugings with non-vanishing de Roo-Wagemans phases
the higher dimensional origin is not yet known.
The compactifications that yield these gaugings might be of unconventional type \cite{Hull:2004in,Dabholkar:2005ve}.
The unifying scheme
presented in this paper should be a useful tool when tackling these questions in a covariant form.
On the other hand, we have so far only presented the gauged theories and have shown their consistency.
It would be interesting to further study these theories
by classifying their ground states, computing the mass spectrum, analyzing stability, etc.

\section*{Acknowledgments}

We are grateful to Henning Samtleben for suggesting this project,
his valuable insights during all stages of this work and for carefully reading the manuscript.
We thank Nikolaos Prezas, Giovanni Villadoro, and Fabio Zwirner for pointing out a sign mistake
in the original version of this paper. All errors remain our own.
This work is partly supported by the EU contracts MRTN-CT-2004-503369
and MRTN-CT-2004- 512194, the DFG grant SA 1336/1-1 and the DAAD -- The German Academic Exchange Service.

\begin{appendix}

\section{Gauged half-maximal supergravities in $D=3$}
\setcounter{equation}{0}
\label{app:D3}

The general gauged half-maximal supergravity in $D=3$ was given in \cite{deWit:2003ja,Nicolai:2001ac}.
Here we shortly describe the underlying group theory and the tensors that parameterize the gauging.
We then give the fermion shift matrices and the scalar potential in the same form as we did in
four and five dimensions. Finally we describe the embedding of the four dimensional gaugings into the 
three dimensional ones. This relation is necessary in order to
calculate the four and five dimensional scalar potentials from the known three dimensional one.

\subsection{General gauging, scalar potential, fermion shift matrices}
\label{app:D3embedding}

The global symmetry group of the ungauged theory is $G={\rm SO}(8,n)$, where
$n$ again counts the number of vector multiplets. The vector fields  $A_\mua{}^{\Ma\Mb}=A_\mua{}^{[\Ma\Mb]}$
transform in the adjoint representation of $G$. Here $\Ma,\Mb=1,\ldots, 8+n$ are ${\rm SO}(8,n)$ vector indices.
The general gauging is parameterized by the two real tensors
$\lambda_{\Ma\Mb\Mc\Md}=\lambda_{[\Ma\Mb\Mc\Md]}$ and
$\lambda_{\Ma\Mb}=\lambda_{(\Ma\Mb)}$, with $\eta^{\Ma\Mb} \lambda_{\Ma\Mb} = 0$,
and one real scalar $\lambda$. Together they constitute the embedding tensor
\begin{align}
   \Theta_{\Ma\Mb\Mc\Md} &= \lambda_{\Ma\Mb\Mc\Md} + \lambda_{[\Mc[\Ma} \, \eta_{\Mb]\Md]} 
                             + \lambda \, \eta_{\Mc[\Ma} \, \eta_{\Mb]\Md} \; ,
\end{align}
which enters into the covariant derivative
\begin{align}
   D_\mua &= \partial_\mua - A_\mua{}^{\Ma\Mb} \Theta_{\Ma\Mb}{}^{\Mc\Md} t_{\Mc\Md} \;.
\end{align}
Due to the above definition the embedding tensor automatically satisfies the linear constraint
\begin{align}
   \Theta_{\Ma\Mb \, \Mc\Md} &= \Theta_{\Mc\Md \, \Ma\Mb} \; .
\end{align}
In addition it has to satisfy the quadratic constraint
\begin{align}
   \Theta_{\Ma\Mb\Mg}{}^{\Mi} \Theta_{\Mc\Md\Mi}{}^{\Mh} - \Theta_{\Mc\Md\Mg}{}^{\Mi} \Theta_{\Ma\Mb\Mi}{}^{\Mh}
      &= \Theta_{\Ma\Mb[\Mc}{}^{\Mi} \Theta_{\Md]\Mi\Mg}{}^{\Mh} \; ,
   \label{D3QC}      
\end{align}
which may be written as a constraint on $\lambda_{\Ma\Mb\Mc\Md}$, $\lambda_{\Ma\Mb}$ and $\lambda$.

The scalars of the theory form the coset ${\rm SO}(8,n)/{\rm SO}(8) \times {\rm SO}(n)$ and in the following we
use the same conventions and notations as for the ${\rm SO}(6,n)/{\rm SO}(6) \times {\rm SO}(n)$ coset in four dimension,
in particular we again have
\begin{align}
   M_{\Ma\Mb} &= {{\cal V}_\Ma}^\xa {{\cal V}_\Mb}^\xa + {{\cal V}_\Ma}^{\ya} {{\cal V}_\Mb}^{\ya} \; , &
   \eta_{\Ma\Mb} &= {{\cal V}_\Ma}^\xa {{\cal V}_\Mb}^\xa - {{\cal V}_\Ma}^{\ya} {{\cal V}_\Mb}^{\ya} \; ,
\end{align}
where now $\xa = 1,\ldots,n$ and $\ya = 1,\ldots, 8$.
In addition we need the scalar dependent object
\begin{align}
   M_{\Ma\Mb\Mc\Md\Me\Mf\Mg\Mh} &= \epsilon_{\ya\yb\yc\yd\ye\yf\yg\yh}
                             {{\cal V}_\Ma}^{\ya} {{\cal V}_\Mb}^{\yb} {{\cal V}_\Mc}^{\yc} 
                             {{\cal V}_\Md}^{\yd} {{\cal V}_\Me}^{\ye} {{\cal V}_\Mf}^{\yf} 
			     {{\cal V}_\Mg}^{\yg} {{\cal V}_\Mh}^{\yh} \; .
\end{align}
The scalar potential then takes the form
\begin{align}
   V &= - \, \frac 1 {24} \, \Bigg[ 
               \lambda_{\Ma\Mb\Mc\Md} \lambda_{\Me\Mf\Mg\Mh} 
             \Big( - \ft 1 2 M^{\Ma\Me} M^{\Mb\Mf} M^{\Mc\Mg} M^{\Md\Mh}
	           + 3 M^{\Ma\Me} M^{\Mb\Mf} \eta^{\Mc\Mg} \eta^{\Md\Mh}
	           \nonumber \\ & \qquad
		   - 4  M^{\Ma\Me} \eta^{\Mb\Mf} \eta^{\Mc\Mg} \eta^{\Md\Mh}
		   + \ft 3 2 M^{\Ma\Me} \eta^{\Mb\Mf} \eta^{\Mc\Mg} \eta^{\Md\Mh}
                   + \ft 1 3 M^{\Ma\Mb\Mc\Md\Me\Mf\Mg\Mh} 
	     \Big)
	\nonumber \\ & \qquad     
           + \lambda_{\Ma\Mb} \lambda_{\Mc\Md}
	     \left( - \ft 3 2 M^{\Ma\Mc} M^{\Mb\Md} + \ft 3 2 \eta^{\Ma\Mc} \eta^{\Mb\Md}
	            + \ft 3 4 M^{\Ma\Mb} M^{\Mc\Md} \right) 
        \nonumber \\ & \qquad		    
	   + 192 \lambda^2 - 24 \, \lambda \, \lambda_{\Ma\Mb} M^{\Ma\Mb} 
	   \Bigg] \; .
    \label{VD3}	   
\end{align}
Although written differently, this is the same potential as given in \cite{deWit:2003ja}.

The maximal compact subgroup of $G$ is $H={\rm SO}(8) \times {\rm SO}(n)$.
All the fermions and the fermion shift matrices $A_1$ and $A_2$ transform under $H$.
Let $A, \, \dot A = 1, \ldots, 8$ be (conjugate) ${\rm SO}(8)$ spinor indices. The Gamma-matrices of ${\rm SO}(8)$
satisfy
\begin{align}
   \Gamma^{(\ya}_{A \dot A} \Gamma^{\yb)}_{B \dot A} &= \delta^{\ya\yb} \delta_{AB} \, ,&
   \Gamma^{\ya\yb}_{AB} & \equiv \Gamma^{[\ya}_{A \dot A} \Gamma^{\yb]}_{B \dot A} \, .
\end{align}
Then the fermion shift matrices $A_1$ and $A_2$ are defined through the so called $T$-tensor as follows \cite{deWit:2003ja}
\begin{align}
   T^{AB \, CD} &= \frac 1 {16} \Gamma^{AB}_{\ya\yb} \Gamma^{CD}_{\yc\yd}
                    {{\cal V}_\Ma}^\ya {{\cal V}_\Mb}^\yb {{\cal V}_\Mc}^\yc {{\cal V}_\Md}^\yd
                    \Theta^{\Ma\Mb \, \Mc\Md} \; , \nonumber \\
   T^{AB \, \ya\xa} &= \frac 1 4 \Gamma^{AB}_{\yc\yd}
                    {{\cal V}_\Ma}^\yc {{\cal V}_\Mb}^\yd {{\cal V}_\Mc}^\ya {{\cal V}_\Md}^\xa
                    \Theta^{\Ma\Mb \, \Mc\Md} \; , \nonumber \\
   A_1^{AB} &= - \frac 8 3 T^{AC \, BC} + \frac 4 {21} \delta^{AB} T^{CD \, CD}  \; , \nonumber \\
   {A_2^{AB}}_{\ya\xa} &= 2 {T^{AB}}_{\ya\xa} 
                        - \frac 2 3 \Gamma^{C(A}_{\ya\yb} {T^{B)C}}_{\yb\xa}
			- \frac 1 {21} \delta^{AB} \Gamma^{CD}_{\ya\yb} {T^{CD}}_{\yb\xa} \; .
\end{align}
The quadratic constraint \eqref{D3QC} guarantees that $A_1$ and $A_2$ satisfy
\begin{align}
    A_1^{AC} A_1^{BC} - {A_2^{AC}}_{\ya\xa} {A_2^{BC}}_{\ya\xa} &= - \frac 1 {128} \delta^{AB} V \; ,
\end{align}
with the scalar potential $V$ appearing on the right hand side.

\subsection{From $D=4$ to $D=3$}
\label{sec:reduction43}

Performing a circle reduction of four dimensional $N=4$ supergravity with $n$ vector multiplets
yields a three dimensional $N=8$ supergravity with $n+2$ vector multiplets. 
The embedding of the global symmetry groups is given by 
\begin{align}
   {\rm SO}(8,n+2) \, \supset \, {\rm SO}(2,2)  \times {\rm SO}(6,n)
                   \, \supset \, {\rm SL}(2) \times {\rm SO}(6,n) \; ,
\end{align}   
where the ${\rm SL}(2)$ is just one of the factors in ${\rm SO}(2,2) = {\rm SL}(2) \times {\rm SL}(2)$.
Accordingly we split the fundamental representation of ${\rm SO}(8,n+2)$ as 
$v^{\tilde \Ma} = (v^\Ma, v^{x \aa})$ where $\aa=1,2$ and $x=1,2$.
Note that the ${\rm SO}(8,n+2)$ vector index is denoted by $\tilde \Ma$, while $\Ma$ is an ${\rm SO}(6,n)$ vector index.
The ${\rm SO}(2,2)$ metric is given by
\begin{align}
   \eta_{x \aa \; y \ab} &= \epsilon_{xy} \epsilon_{\aa\ab} \, , &
   \text{which yields} &&
    \eta_{x \aa \; y \ab} \eta^{y \ab \; z \ac} &= \delta^{z \ac}_{x \aa} \; .
\end{align}
The ${\rm SL}(2)$ generators $t_{(\aa\ab)}$, $t_{(xy)}$ and the ${\rm SO}(2,2)$
generators $t_{x \aa \, y \ab}=t_{y \ab\, x\aa}$
are related as follows
\begin{align}
   t_{x \aa \, y \ab} = - \frac 1 2 \left( \epsilon_{\aa\ab} t_{xy} + \epsilon_{xy} t_{\aa\ab} \right) \; ,
\end{align}
where we use the conventions $(t_{\cMa\cMb})_\cMc{}^\cMd = \delta^\cMd_{[\cMa} \eta^{\phantom{\cMd}}_{\cMb]\cMc}$ for the
${\rm SO}(2,2)$ generators ($\cMa=x\aa$).
The embedding of the $D=4$ vector fields into the $D=3$ ones is then given by
\begin{align}
   A_\mua^{\Ma \alpha} \, = \, A_\mua^{\Ma \, 1 \alpha} \; ,
\end{align}
where $A_\mua^{\Ma \, 1 \alpha}$ denotes the corresponding components of the $D=3$ vector fields
$A_\mua^{\tilde \Ma \tilde \Mb} = A_\mua^{[\tilde \Ma \tilde \Mb]}$.
Analogous to the reduction from $D=5$ to $D=4$ described in section \ref{sec:reduction},
now the covariant derivatives in $D=4$ and $D=3$ have to agree for those terms already present in $D=4$, i.e.
\begin{align}
   D_\mua  \, &\supset \, \partial_\mua 
       - 2 A_\mua^{\Ma \, 1 \alpha} {\Theta_{\Ma \, 1 \alpha}}^{\Mb\Mc} t_{\Mb\Mc}
       + A_\mua^{\Ma \, 1 \alpha} {\Theta_{\Ma \, 1 \alpha}}^{x \ab \, y \ac} \, \epsilon_{xy} \, t_{\ab\ac}
       \nonumber \\
        &=  \, \partial_\mua - A_\mua{}^{\Ma\aa} {\Theta_{\Ma\aa}}^{\Mb\Mc} t_{\Mb\Mc}
                             - A_\mua{}^{\Ma\aa} {\Theta_{\Ma\aa}}^{\ab\ac} t_{\ab\ac}  \; .    
\end{align}
This yields
\begin{align}
   \lambda_{1 \aa \, \Ma\Mb\Mc} &= - \, \ft 1 2 \, f_{\aa \Ma\Mb\Mc} \; , &
   \lambda_{\Ma \, 1\aa \, x\ab \, y\ac} \epsilon^{xy} &= \, \ft 1 2 \, \epsilon_{\aa(\ac} \xi_{\ab)\Ma} \; , &
   \lambda_{1 \aa \, \Ma} &= \xi_{\aa \Ma} \; ,
\end{align}
while we demand the other components of $\lambda_{\tilde \Ma \tilde \Mb \tilde \Mc \tilde \Md}$ and
$\lambda_{\tilde \Ma \tilde \Mb}$ to vanish and also $\lambda=0$. However,
the antisymmetry of $\lambda_{\tilde \Ma \tilde \Mb \tilde \Mc \tilde \Md}$ 
and the symmetry of $\lambda_{\tilde \Ma \tilde \Mb}$ has to be imposed, for example
\begin{align}
   \lambda_{\Ma \, z\aa \, x\ab \, y\ac} &= \tilde \lambda_{\Ma \, [\{z\aa\} \, \{x\ab\} \, \{y\ac\}] } \, , &
   \tilde \lambda_{\Ma \, z\aa \, x\ab \, y\ac} &= \, \ft 1 2 \, \delta_z^1 \epsilon_{xy} \epsilon_{\aa(\ac} \xi_{\ab)\Ma} \, .
\end{align}
We have thus defined the embedding of the four dimensional gaugings into the three dimensional ones.
The quadratic constraint \eqref{D3QC} in $D=3$ is satisfied iff the $D=4$ quadratic constraint \eqref{QConD4}
is satisfied. The $D=3$ scalar potential \eqref{VD3} reduces to the $D=4$ potential \eqref{VD4} when all $D=3$
extra scalars are set to the origin.

\end{appendix}


\providecommand{\href}[2]{#2}\begingroup\raggedright\endgroup

\end{document}